\documentclass[a4paper,11pt]{article}

\pdfoutput=1 

\usepackage{fullpage}

\usepackage{mathrsfs}
\usepackage{comment}
\usepackage{centernot}
\usepackage{mathtools}
\usepackage{stmaryrd}

\usepackage{latexsym,amssymb,amstext,amsmath}
\usepackage{hyperref}
\usepackage{graphbox,graphicx}
\usepackage{subcaption}
\usepackage{amsthm}
\usepackage{esint}
\usepackage{float}

\def\be {\begin{equation}}
\def\ee {\end{equation}}
\def\bea {\begin{eqnarray}}
\def\eea {\end{eqnarray}}

\def\beq{\begin{equation}}
\def\eeq{\end{equation}}
\def\beqa{\begin{eqnarray}}
\def\eeqa{\end{eqnarray}}

\newtheorem{theorem}{Theorem}[section]

\newtheorem{proposition}[theorem]{Proposition}
\newtheorem{corollary}[theorem]{Corollary}

\theoremstyle{definition}

\newtheorem{remark}[theorem]{Remark}
\newtheorem{definition}[theorem]{Definition}

\DeclareMathOperator{\Tr}{Tr}

\begin{document}

\title{\textbf{\LARGE  
Generating new gravitational solutions \\ by matrix multiplication
}}
\author{M.~Cristina C\^amara and Gabriel Lopes Cardoso}
\date{\small 
\vspace{-5ex}
\begin{quote}
\emph{
\begin{itemize}
\item[]
Center for Mathematical Analysis, Geometry and Dynamical Systems,\\
  Department of Mathematics, 
  Instituto Superior T\'ecnico, Universidade de Lisboa,\\
  Av. Rovisco Pais, 1049-001 Lisboa, Portugal
  \end{itemize}
}
\end{quote}
{\tt 
cristina.camara@tecnico.ulisboa.pt, gabriel.lopes.cardoso@tecnico.ulisboa.pt}
}
\maketitle

\begin{abstract}
\noindent
Explicit solutions to the non-linear field equations of some gravitational theories can be obtained, by means of a Riemann-Hilbert approach, from a canonical Wiener-Hopf factorisation of certain matrix functions called monodromy matrices. In this paper we describe other types of factorisation from which solutions can be constructed in a similar way. Our approach is based on an invariance problem, which does not constitute a Riemann-Hilbert problem and allows to construct solutions that could not have been obtained by Wiener-Hopf factorisation of a monodromy matrix. It gives rise to a novel solution generating method based on matrix multiplications. We show, in particular, that new solutions can be obtained by multiplicative deformation of the canonical Wiener-Hopf factorisation, provided the latter exists, and that one can superpose such solutions. Examples of applications include Kasner, Einstein-Rosen wave and gravitational pulse wave solutions.
\\

%
%
\end{abstract}

\section{Introduction  }

Finding exact solutions to the Einstein field equations in space-time dimensions $D \geq 4$  is, in general, a difficult task. They can, however, be obtained under simplifying assumptions. In this paper we focus on solutions that only depend on two of the $D$  space-time coordinates, denoted here by $(\rho,v)$,
in which case the field equations reduce to non-linear PDE's in two dimensions. 

This procedure applies not only to General Relativity,
but also to various other gravity and supergravity theories in $D \geq 4$ space-time dimensions, as is well-known \cite{Breitenlohner:1987dg}. The dimensional reduction to two dimensions, called Ehlers reduction, is performed in two steps.
The first step requires the presence of $D-3$ commuting Killing symmetries, in which case these theories can be reduced to three space-time dimensions and
reformulated as theories describing the coupling of non-linear sigma-models based on coset spaces $G/H$ to three-dimensional gravity. A further reduction
on another commuting Killing symmetry gives rise to field equations in two dimensions that are integrable \cite{Belinsky:1971nt,Breitenlohner:1986um,Nicolai:1991tt,Korotkin:2023lrg}.
An early example thereof is provided by the Ernst equation that arises when dimensionally reducing Einstein's vacuum field equations \cite{Ernst:1967wx} and the coupled
Einstein-Maxwell field equations \cite{Ernst:1967by}.  

The resulting PDE's in two dimensions are an integrable system, i.e. they appear as a compatibility condition for an auxiliary linear system, called a Lax pair \cite{Its}.
There are two descriptions of such a Lax pair, which can be shown to be equivalent \cite{Figueras:2009mc}, but they originate rather different approaches to solving the 
field equations.

One is called the Belinski-Zakharov 
linear system \cite{Belinsky:1971nt}. It is particularly well suited to obtain exact solutions to the field equations if one focusses on solitonic
solutions which can be obtained by the inverse scattering method. The latter was discovered by Gardner, Greene, Kruskal and Miura for solving the KdV equation
\cite{Gardner:1967wc,Ablowitz:1991xb} and has since then been developed and expanded to a wide class of new integrable systems. The inverse scattering method, 
which was applied in \cite{Belinsky:1971nt} to obtain exact solutions to the Einstein field
equations and was subsequently generalized to five-dimensional minimal supergravity \cite{Figueras:2009mc}, is  a systematic procedure for generating new solutions from a 
previously known solution, called a seed solution.
If $\psi_0$ is a seed solution, one looks for solutions to the Lax pair
of the form $\chi \, \psi_0$, where $\chi$ is called the dressing matrix, by assuming a particular rational form for $\chi$, involving only simple poles, and substituting
in the Lax pair.
This approach, however, presents some difficulties in the current context; namely, the new solution is not, in general, an element of the coset $G/H$ (see \cite{Figueras:2009mc} for more details).

The other Lax pair, called the Breitenlohner-Maison linear system \cite{Breitenlohner:1986um}, adopts a group theoretic approach by taking into account the group structure
of the dimensionally reduced model and
makes crucial use of an involution $\natural$ associated with the coset $G/H$. It relies heavily on constructing solutions to a Riemann-Hilbert factorisation problem
for a so-called monodromy matrix. Indeed, the Breitenlohner-Maison linear system depends on a complex parameter $\tau$, which is allowed to vary in a certain algebraic curve, whose presence is the fundamental ingredient that allows for reformulating the problem of solving the PDE's in two dimensions in terms of Riemann-Hilbert problems and Wiener-Hopf factorisation of matrix functions, with respect to a contour in the complex $\tau$-plane.

In this paper we will use the approach based on the Breitenlohner-Maison linear system for explicitly determining solutions to the Einstein field equations.
This approach was 
introduced in \cite{Breitenlohner:1986um,Nicolai:1991tt} and has recently become the object of rekindled interest,
see for instance \cite{Aniceto:2019rhg, Camara:2017hez,Cardoso:2017cgi,Chakrabarty:2014ora,Gohberg2003FactorizationAI,Katsimpouri:2012ky,Penna:2021kua} and references therein. The associated Riemann-Hilbert factorisation problem 
is, in general, 
a difficult one to solve, since that factorisation is an analytic tool whose implementation is not at all algorithmic and often requires developing 
"custom-made" methods of solution \cite{Its}. However, one can take advantage of the recent progress in explicit factorisation methods which enables us to solve the factorisation problem for a large class of matrices, including all rational matrix functions (whatever the order of their poles), as well as 
various classes of non-rational ones (see for example \cite{AAM,CLS,CAFSPFS,CAFSPFS2,Cardoso:2017cgi,CG,LS,KisilAMR}).

There are several types of Wiener-Hopf factorisation. One such type is called canonical Wiener-Hopf factorisation, 
defined in Section 2b. This type of factorisation is the one 
that has been the main focus of the works  \cite{Aniceto:2019rhg, Camara:2017hez, Katsimpouri:2012ky}.
It was proved in {\cite [Theorem 6.1]{Aniceto:2019rhg}}
 that if a given monodromy matrix ${\cal M}_{\rho, v} (\tau) $ (see \eqref{mont})
possesses a canonical Wiener-Hopf factorisation 
with respect to a certain contour $\Gamma$ (see \eqref{whc}), 
\bea
{\cal M}_{\rho, v} (\tau) = {\cal M}^-_{\rho, v} (\tau) X (\tau, \rho, v) \quad \text{on} \quad \Gamma, 
\label{whc1}
\eea
then it yields a matrix $M(\rho,v)$, obtained by taking the limit of $ {\cal M}^-_{\rho, v} (\tau)$ when $\tau \rightarrow \infty$, which provides a solution to the
field equations in $D$ dimensions.

In this paper we take the Breitenlohner-Maison Lax pair as a starting point and significantly extend the scope of application of the Wiener-Hopf factorisation techniques to solve that linear system. Our approach, although motivated by the Breitenlohner-Maison factorisation method, will be based on the solutions of what we call a $\tau$-invariance problem,
which does not constitute a Riemann-Hilbert problem, and can be seen as generalising the Breitenlohner-Maison approach. 
This invariance problem constitutes one of the novel aspects of this paper.
We use it to propose a new solution generating method and to study several questions which arise naturally in the context of factorisation theory, as follows.

Question 1: Can we find other types of factorisation of a monodromy matrix that, in a similar manner, yield
solutions to the field equations? Starting from the simplest possible example, take the identity monodromy matrix, whose canonical Wiener-Hopf factorisation is trivial.
Can we find another factorisation of the identity matrix
with less restrictive conditions on the factors, providing a nontrivial solution to the field equations?

Question 2: Given two solutions $M_1(\rho, v)$ and $M_2(\rho, v)$, can we multiply them to obtain a new solution $(M_1 M_2) (\rho, v)$ which may then
be interpreted as resulting from $M_1(\rho, v)$ by ``multiplicative deformation", via multiplication by $M_2(\rho, v)$?

Question 3: In the context of Question 2, suppose that we have an infinite family of solutions, indexed by a continuous parameter $k \in \mathbb{R}^+$ 
(that can be seen as representing a wave number), which constitutes an Abelian group of solutions. Then a finite product, seen as a superposition of
a finite number of waves, is also a solution to the field equations. Can we extend this result to the superposition of an infinite number of such solutions?


The paper is organised as follows. To keep it as self-contained as possible, in Section  \ref{sec:backg}
we present some background material. First we briefly review the description of the dimensionally
reduced gravitational  field equations as an integrable system associated with a Lax pair called  Breitenlohner-Maison linear system. Secondly we
describe how to solve the field equations by means of canonical bounded Wiener-Hopf factorisations. Finally we briefly review some known results about factorisation in the algebra of H\"older continuous functions.
In Section \ref{sec:nsfd} we address Questions 1 and 2. We introduce 
 the $\tau$-invariance problem (Theorem \ref{theo3.3})
and use it to describe a solution generating method based on matrix multiplications,
 see Theorem \ref{theoMN}. Moreover, we describe a class of solutions to the $\tau$-invariance problem, which also satisfy the field equations, and use it
to obtain a class of multiplicative deformations providing  
  an application of the solution generating method.
In Section \ref{sec:cosmk} we apply the previous results to cosmological Kasner solutions in four space-time dimensions.
We also briefly discuss conserved currents. In Section \ref{E-R} we use the results of Section \ref{sec:nsfd} to establish a remarkable
property of 
the Einstein-Rosen wave solutions in General Relativity, and we use it to construct other solutions which would be difficult to obtain by directly solving
the non-linear field equations.
In Section \ref{sec:supER} we address Question 3 by discussing it in the specific context of the superposition of infinitely many Einstein-Rosen wave solutions.
In Section \ref{sec:AX} we use the previous results
to infer some interesting relations between the matrix one-form $A = M^{-1}dM$ and the matrix function $X(\tau,\rho,v)$ in a canonical Wiener-Hopf factorisation 
\eqref{whc1}.


\section{Notation and background results} \label{sec:backg}

\subsection{The field equations as an integrable system: Lax pair and spectral curve \label{lsc}}

The field equations of gravitational theories in $D$ space-time dimensions are a system of non-linear PDE's for the space-time metric (and other fields) for which
obtaining exact solutions is, in general, a difficult task. Exact solutions can however be obtained under simplifying assumptions, such as imposing spherical
symmetry. An example of such a solution, which is of great physical and mathematical interest, is the well known Schwarzschild solution.

Here we focus on solutions that only depend on two of the $D$  space-time coordinates and which 
result from performing a two-step dimensional reduction 
of the original theory. 
Below we summarize the resulting equations \cite{Lu:2007jc,Nicolai:1991tt,Schwarz:1995af}.

After performing this two-step dimensional reduction, the field equations become field equations in two dimensions, 
\bea 
d \left( \rho \star A \right) = 0 \;\;\;,\;\;\; \text{with} \; \; A = M^{-1} dM  \;,
\label{fi2d}
\eea
where $M \in G/H$ is a coset representative of the symmetric space $G/H$ that arises in the two-step reduction.
$G/H$ is invariant under an involution called generalised transposition, which we
denote by $\natural$, and hence $M^{\natural} = M $. The field equation \eqref{fi2d}
is a matricial non-linear second order PDE depending on two coordinates, which we take to be $\rho > 0$ and $v \in \mathbb{R}$, called {\it Weyl coordinates}. Accordingly, we will denote the $(\rho,v)$ upper-half plane by Weyl upper-half plane.
In \eqref{fi2d}, 
$\star$ denotes the Hodge star operator in two dimensions, satisfying
\bea
\star d \rho = - \lambda \, dv \;\;\;,\;\;\; \star dv = d \rho \;\;\;,\;\;\; (\star)^2 = - \lambda \, {\rm id} \;,
\label{stho}
\eea
where $\lambda = \pm 1$, depending on the form of the two-dimensional line element $ds_2^2$ (given by $ ds_2^2 = \sigma d\rho^2 + \varepsilon dv^2, \,
\lambda = \sigma \varepsilon$).

We only consider here cases where a solution to the field equations in $D$ space-time dimensions is uniquely determined by 
a solution $M(\rho,v)$ to \eqref{fi2d}, see for example Appendix A of \cite{Camara:2017hez}. In particular, 
when $D=4$, the associated four-dimensional space-time metric takes the 
(Weyl-Lewis-Papapetrou) form,
\bea
ds_4^2 = - \lambda \, \Delta (dy  + B  d \phi  )^2 + \Delta^{-1} 
\left(  e^\psi \, ds_2^2 + \rho^2 d\phi^2 \right) \;,
\label{4dWLP}
\eea
where $\Delta, B$ are functions of $(\rho, v)$ uniquely determined  by the solution $M(\rho,v)$ of \eqref{fi2d} 
(see \eqref{M22} for the case of 
$2 \times 2$ matrices $M(\rho, v)$) and $\psi(\rho,v)$ is a scalar function
determined  by integration \cite{Lu:2007jc,Schwarz:1995af} from
\be
\partial_\rho \psi = \tfrac{1}{4} \, \rho \,  \Tr \left(A_\rho^2 - \lambda \, A_v^2\right) \,,\qquad \partial_v \psi = \tfrac{1}{2} \, \rho \, 
 \Tr \left(A_\rho A_v \right) \,. 
 \label{eq_psi}
\ee
Note that, in \eqref{eq_psi}, 
$\partial_v \left( \partial_\rho \psi \right) = \partial_\rho \left( \partial_v \psi \right)$, as can be verified by using \eqref{fi2d}, \eqref{stho} and the
property $dA + A\wedge A =0$.
For examples of Weyl metrics in dimensions $D > 4$ see \cite{Emparan:2001wk}.

{
\remark
\label {r2.1}
Since the class of solutions to the field equations in $D$ space-time dimensions that we study here is in a one-to-one correspondence with the class of
solutions 
$M(\rho,v)$ satisfying \eqref{fi2d}, we identify the two classes and henceforth we refer to $M(\rho,v)$ as  a solution to the field equations.
\\
}

The non-linear field equation \eqref{fi2d} is an integrable system \cite{Its}, since it can be viewed as the compatibility condition for an auxiliary
linear system of differential equations (a Lax pair) depending on an auxiliary complex parameter $\tau$, which is allowed
to vary in an algebraic curve depending on $\rho$ and $v$. The Lax pair for \eqref{fi2d} takes the form \cite{Lu:2007jc}
\begin{equation}
	\tau \left( dX + A X \right) = \star \, dX \;.
	\label{lax}
\end{equation}
This linear system (for the unknown matrix $X(\tau, \rho, v)$) will be called the Breitenlohner-Maison linear system \cite{Breitenlohner:1986um}.
The complex parameter $\tau$ plays a crucial role here in allowing to bring the tools of complex analysis into the problem, thereby transforming the original
problem of solving the  non-linear field equation \eqref{fi2d} into a Riemann-Hilbert problem, described in Section 2b.

In the case of \eqref {lax}, 
the complex parameter $\tau$ (called the {\it spectral parameter}) is not a constant, but must satisfy
an algebraic relation involving another complex variable $\omega$ 
as well as
the two space-time coordinates $\rho$ and $v$,
\bea
\omega = v + \frac{\lambda}{2}    \, \rho \, \frac{\lambda -   \tau^2}{\tau} \;\;\;,\;\;\; \tau \in \mathbb{C} \backslash \{0\} \;.
\label{spec}
\eea
This relation is invariant under the involution
\bea
\tau \xmapsto[\, i_{\lambda}]{}
- \frac{\lambda}{\tau}  \;\; \text{for all} \;\; \tau \neq 0\;.
\label{invt}
\eea
Note that, for any given $\omega$, the spectral relation \eqref{spec} is satisfied if we replace $\tau$ by\footnote{Note 
that the expression for $\rho \, \varphi_{\omega}$ is reminiscent of the expression that arises in the context of rod structures in General Relativity \cite{Emparan:2001wk}.}
\bea
\varphi_{\omega} (\rho, v)= \frac{-\lambda  (\omega-v) + \sqrt{(\omega-v)^2+\lambda \rho^2} }{\rho} 
\label{vpp}
\eea 
or
\bea
{\tilde \varphi}_{\omega} (\rho, v) = \frac{-\lambda  (\omega-v) - \sqrt{(\omega-v)^2+\lambda \rho^2} }{\rho} = - \frac{\lambda}{\varphi_{\omega} (\rho, v)}\;.
\label{vpp2}
\eea 
Then, defining the set ${\cal T}$ to be the set of functions of the form \eqref{vpp} or \eqref{vpp2}, 
by studying the compatibility condition for the Breitenlohner-Maison linear system \cite{Aniceto:2019rhg, Camara:2017hez, Lu:2007jc} one establishes the following. 

{\theorem 
\cite{Lu:2007jc}
{\cite [Theorem 4.2]{Aniceto:2019rhg}}
Let $\varphi \in {\cal T}$. If there exists $X(\tau, \rho, v)$ such that $X$ is twice continuously differentiable and invertible with $X^{-1}$ continuously 
differentiable with respect to $(\rho, v)$ and, for $\tau=\varphi$, 
\bea
\tau \left( dX + A X \right) = \star \, dX \;, \quad \text{with} \quad A = M^{-1} d M, 
\label{varXA}
\eea
then $M$ is a solution to the field equation $d \left( \rho \star A \right) = 0, \; A = M^{-1} d M$.\\
}

{
\remark\label {r2.2}

Since $\omega \in \mathbb C$ and $\varphi \in {\cal T}$ is a complex valued function, when we refer to 
 $A(\rho,v)$ (and $M(\rho,v)$) as a solution of the field equation \eqref{fi2d},
 we take them as being complex valued. One may seek a real or a complex solution $M(\rho,v)$. In the former case, restrictions  may have to be imposed on the domain of $(\rho,v)$.
 \\
 }

Note that, although the Breitenlohner-Maison linear system \eqref{varXA} is a system of linear PDE's for the unknown $X$ with coefficient $A(\rho, v)$, one must
in fact look at it as a system of  equations for the pair of unknowns $(X,A)$ if the field equations are to be solved; indeed, in general $A$ is not given and 
obtaining $A = M^{-1} d M$ such that
\eqref{fi2d} holds is precisely our goal.

One way to address this question, providing at the same time both a solution $M(\rho, v)$ to the field equation \eqref{fi2d} and a solution $X$ 
to the Lax pair \eqref{varXA} with input  $A = M^{-1} d M$, is given by the canonical Wiener-Hopf factorisation (also known as Birkhoff factorisation) of
a so-called monodromy matrix. Obtaining this factorisation is equivalent to solving a particular kind of Riemann-Hilbert problems, as described in the next section.

\subsection{{Solving the field equation by canonical Wiener-Hopf factorisation \label{sec:RHP}}}

Let 
$\Gamma$ be a simple closed contour in $\mathbb{C}$, encircling the origin,
and let $\mathbb{D}^+_{\Gamma}$ and $\mathbb{D}^-_{\Gamma}$ denote the interior and the exterior of $\Gamma$  (including the point $\infty$), respectively.

{\it A Riemann-Hilbert problem} is the problem of determining a function $\phi$, analytic in $\mathbb{C} \backslash \Gamma$, satisfying a jump condition
across $\Gamma$.

 A particularly important Riemann-Hilbert problem is the so-called Wiener-Hopf factorisation problem \cite{CDR,MP}. Developing efficient methods to obtain such a factorisation has recently attracted renewed interest due to its importance in applications \cite{AAM, GKR, PRD}. We look here for a canonical bounded Wiener-Hopf factorisation (abbreviated to {\it canonical Wiener-Hopf factorisation}) \cite{CDR,LS,MP},  where one seeks $n \times n$ matrix functions ${\cal M}_+$
 and ${\cal M}_-$  which are analytic and bounded in $\mathbb{D}^+_{\Gamma}$ and $\mathbb{D}^-_{\Gamma}$  (respectively), as well as their  
 inverses, and such that for a given $n \times n$ matrix function ${\cal M}$
 defined on $\Gamma$ one has  
  \bea
{\cal M}(\tau) = {\cal M}_- (\tau)  \, {\cal M}_+ (\tau) \;\; \text{on} \;\; \Gamma \;,
\label{GGG}
\eea
where ${\cal M}_-$ and ${\cal M}_+$ are identified with their (non-tangential) boundary value functions on $\Gamma$.

When such a factorisation exists, the factors ${\cal M}_{\pm}$  are unique up to constant factors. One can then impose a normalisation
condition ${\cal M}_+ (0) = \mathbb{I}_{n \times n}$ which uniquely determine the factors ${\cal M}_{\pm}$ in \eqref{GGG}.

{\remark

Note that the same canonical Wiener-Hopf factorisation of a monodromy matrix can be obtained from different contours, as it
happens, for instance, in Section \ref{E-R} (see Remark \ref{conhom}).
\\
}

We will assume here that the elements of ${\cal M}$ are H\"older continuous functions with exponent $\mu \in \, ]0,1[$ on $\Gamma$
and $\det {\cal M} \neq 0$ admits a (scalar) factorisation of the form \eqref{GGG}  \cite{CG,MP}; in this case we may assume without loss of generality that
$\det {\cal M} =1$.

Now let ${\cal M} (\omega)$ be a matrix function of the complex variable $\omega$ with ${\cal M} (\omega) = {\cal M}^{\natural}  (\omega)$, 
and denote by ${\cal M}_{\rho, v} (\tau)$ the matrix
that is obtained from  ${\cal M} (\omega)$ by composition using the spectral relation \eqref{spec}, i.e.
\bea
{\cal M}_{\rho, v} (\tau) = {\cal M} \left( v + \frac{\lambda}{2}    \, \rho \, \frac{\lambda -   \tau^2}{\tau}  \right)
\label{mont}
\eea
defined for all $\tau \in \mathbb{C} \backslash \{0\}, v \in \mathbb{R}, \rho \in \mathbb{R}^+$. 
In what follows, we will also use the notation ${\cal M} (\tau, \rho, v)$ which we
identify with ${\cal M}_{\rho, v} (\tau) $.

If ${\cal M}_{\rho, v} (\tau)$ admits a canonical Wiener-Hopf factorisation ${\cal M}_{\rho, v} (\tau) = {\cal M}^-_{\rho, v} (\tau) {\cal M}^+_{\rho, v} (\tau) $ with respect to $\Gamma$
and if we denote by $X (\tau, \rho, v)$ the factor ${\cal M}^+_{\rho, v} (\tau) $ when the latter is normalized to ${\cal M}^+_{\rho, v} (0) = \mathbb{I}$, one
can show that (see Section 3 of \cite{Camara:2017hez})
\bea
{\cal M} (\omega) = {\cal M}^{\natural}  (\omega) \Rightarrow  {\cal M}^-_{\rho, v} (\tau) = X^{\natural} \left( -\frac{\lambda}{\tau} , \rho, v \right) \, M(\rho,v) .
\eea
The canonical Wiener-Hopf factorisation can then be written as 
\bea
{\cal M}_{\rho, v} (\tau) = X^{\natural} \left( -\frac{\lambda}{\tau}, \rho, v \right) \, M(\rho,v) \, X (\tau, \rho, v) \quad \text{on} \quad \Gamma
\label{canfacfac}
\eea
with 
\bea
 X (\tau = 0, \rho, v) = \mathbb{I}
 \label{noI}
 \eea
 and 
 \bea
  M^{\natural} (\rho,v) =  M(\rho,v) .
  \eea
  Analogously, 
  \bea
  \det {\cal M} (\omega) = 1 \Rightarrow \det M(\rho,v) = 1 .
  \eea
  Indeed, we have from ${\cal M}_{\rho, v} (\tau) = {\cal M}^-_{\rho, v} (\tau) X(\tau, \rho, v)$ that
  \bea
  1 = \det \left(  {\cal M}^-_{\rho, v} (\tau) \right) \, \det X(\tau, \rho, v) \Leftrightarrow \det X(\tau, \rho, v) = \left( \det \left(  {\cal M}^-_{\rho, v} (\tau) \right)\right)^{-1} , 
  \eea
  and since the left hand side of the last equality is analytic and bounded in $\mathbb{D}^+_{\Gamma}$ while the right hand side is analytic and bounded in 
  $\mathbb{D}^-_{\Gamma}$, both are constant. The latter determinant must be equal to $1$, since $X(0, \rho, v) = \mathbb{I}$ for all $\rho, v$. 
  Therefore, since $ \det \left(  {\cal M}^-_{\rho, v} (\tau) \right) = 1$,  it follows that
  $\det M(\rho, v) =   \underset{\tau \to \infty} \lim  \det \left(  {\cal M}^-_{\rho, v} (\tau) \right) = 1$.

It was shown in Theorem 6.1 of \cite{Aniceto:2019rhg} that, under 
very general assumptions\footnote{
These assumptions, which may seem
complicated and technical, are in fact intended so as to allow for the broadest possible range of applicability. They are easily
seen to be satisfied in all known physically significant cases.},
the canonical Wiener-Hopf factorisation of ${\cal M}_{\rho, v} (\tau)$ with respect to an appropriate contour $\Gamma$ in the complex
$\tau$-plane, determines a solution to the field equation \eqref{fi2d}. Let us state this theorem. 
We denote here by $\natural$ an involution in the space of matrix functions ${\cal N}(\tau, \rho, v)$, which coincides with
the generalised transposition (also denoted by $\natural$) when ${\cal N}(\tau, \rho, v)$ does not depend on $\tau$ and belongs to
$G/H$, and which commutes with the involution $i_{\lambda}$ defined in \eqref{invt}. We also assume that if 
${\cal N}(\tau, \rho, v)$ is analytic with respect
to $\tau$ in a region $\mathcal{U} \subset \mathbb{C}$, then ${\cal N}^{\natural} (\tau, \rho, v)$ is also analytic for $\tau \in \mathcal{U}$ and $\partial_{\tau} ({\cal N}^{\natural})
= (\partial_{\tau} {\cal N} )^{\natural}$.

{
\theorem
\label{theoraccr}
{\cite [Theorem 6.1]{Aniceto:2019rhg}}
Let the following assumptions hold:

(A.1)  
There exists an open set $S$ such that, for every 
 $(\rho_0, v_0)\in S$, one can find a
simple closed contour $\Gamma$ in the $\tau$-plane, which is 
invariant under $\tau \xmapsto[\, i_{\lambda}]{}
- \frac{\lambda}{\tau} $ and 
encircles the origin, such that the following holds:
\\
for all 
$(\rho,v)$ in a neighbourhood of $(\rho_0, v_0)$, 
the matrix \eqref{mont}, as well as its inverse, is 
analytic in a region $O$ 
in the $\tau$-plane containing 
$\Gamma$. 
We require $O$ to be 
 invariant under  $i_\lambda$, and such that  ${\cal M}^{\natural}_{\rho, v} (\tau)  = {\cal M}_{\rho, v} (\tau) $ on $O$;
 
 (A.2) for any $(\rho, v)$ in a neighbourhood of $(\rho_0, v_0)$, ${\cal M}_{\rho, v} (\tau) $ admits a canonical Wiener-Hopf factorisation
 with respect to $\Gamma$,
\bea
{\cal M}_{\rho, v} (\tau) = {\cal M}^-_{\rho, v} (\tau) X (\tau, \rho, v) \;,
\label{whc}
\eea
where the ``plus" factor $X$ satisfies the normalisation condition
\bea
X (0, \rho, v) = \mathbb{I}_{n \times n} \;\; \text{for all $(\rho,v)$ in a region $S \subset \mathbb{R}^+ \times \mathbb{R}$} \;;
\eea

(A.3) the matrix function $X (\tau, \rho, v)$, for each $\tau \in \mathbb{D}^+_{\Gamma} \cup O$, and 
\bea 
M(\rho,v) : = \displaystyle{\lim_{\tau \rightarrow \infty}} {\cal M}^-_{\rho, v} (\tau) 
\label{defM}
\eea
are of class $C^2$ (with respect to $(\rho, v)$) and $\frac{\partial X}{\partial \rho} $ and  $\frac{\partial X}{\partial v} $
are analytic as functions of $\tau$ in $ \mathbb{D}^+_{\Gamma} \cup O$.

Then $M(\rho, v)$ defined by \eqref{defM} 
is a solution to the field equation \eqref{fi2d} and $ X (\tau, \rho, v) \vert_{\tau = \varphi}$ is a 
solution to the linear system \eqref{varXA}.\\
}

We call ${\cal M}_{\rho, v} (\tau)$ the {\it monodromy matrix} for $M(\rho, v)$. A matrix function $M(\rho, v)$ obtained from the canonical Wiener-Hopf factorisation \eqref{whc} as in the previous theorem  is called a {\it canonical solution} (w.r.t. $\Gamma$) to the field equation \eqref{fi2d}; $X (\tau, \rho, v)$ is the  {\it associated solution} to the linear system \eqref{varXA} (where the substitution of $\tau$ by $\varphi_{\omega} (\rho,v)$ 
 is implicit).

{
\remark

The canonical Wiener-Hopf factorisation \eqref{whc} is obtained by taking $\tau$ as the variable in the complex plane and $(\rho, v)$ as parameters, and choosing, for each  $(\rho, v)$, a certain contour $\Gamma$. Therefore it should be clear that this contour, in the neighborhood of which the  equality \eqref{whc} holds, depends on  $(\rho, v)$. This dependence will be implicit, unless necessary for the exposition, throughout the paper.\\
}

Obtaining conditions for the existence of a canonical Wiener-Hopf factorisation for ${\cal M}$ and explicit expressions for the factors in \eqref{GGG} is a difficult task 
in general \cite{KisilAMR}. However, the following simple but powerful result 
 guarantees that a canonical Wiener-Hopf factorisation exists for all scalar functions ${\cal M}_{\rho, v} (\tau)$:

{\theorem \label{theojoh}
\cite{Cardoso:2017cgi} Any  scalar complex valued function 
\bea
f_{\rho,v} (\tau) = f \left( v + \frac{\lambda}{2}    \, \rho \, \frac{\lambda -   \tau^2}{\tau}  \right)
\eea 
admits a canonical Wiener-Hopf factorisation w.r.t. any closed contour  $\Gamma$, invariant under 
 $\tau \xmapsto[\, i_{\lambda}]{}
- \frac{\lambda}{\tau} $ and encircling the origin, on which it is continuous and non-vanishing. Naturally, the factorisation may depend on the contour.\\
}

The previous theorem implies, in particular, the existence of a canonical Wiener-Hopf factorisation for all monodromy matrices which are reducible, through products
by constant matrices, to triangular form
\cite {LS, MP} and, therefore, for all diagonal matrices. Furthermore, 
as explained in Section \ref{sec:2c},
the canonical Wiener-Hopf factorisation of a scalar function in $ C^{\mu} (\Gamma)$ can be explicitly obtained
via known formulae involving Cauchy-type integrals,  
and therefore every diagonal matrix function
can be explicitly factorised either in terms of known functions or in terms of simple Cauchy-type integrals.
This distinguishes the class of diagonal monodromy matrices. In the next section we explore their unique properties
and illustrate how one can construct highly non-trivial exact solutions to \eqref{fi2d} by using 
simple complex analytic tools.

\vskip 5mm

{\remark 
\label{admc}
In the remainder of this paper, $\Gamma$ will denote a simple, closed contour in the complex $\tau$-plane that encircles the origin and
is invariant under $\tau \xmapsto[\, i_{\lambda}]{}
- \frac{\lambda}{\tau} $, which will be called an {\em admissible contour}, and we assume that $(\rho,v)$ belongs to an open set $S$ as in Theorem \ref{theoraccr}.\\
}

\subsection{Factorisation in the algebra of H\"older continuous functions \label{sec:2c}}

Let $f(\tau)$ be H\"older a continuous function, i.e. satisfying a H\"older condition with exponent $\mu \in \,  ] 0, 1[$, on a simple closed
curve $\Gamma$ encircling $0$, oriented anti-clockwise. We denote by $C^{\mu} (\Gamma)$ the space of all such functions.
For any $f \in C^{\mu} (\Gamma)$ we can define the Cauchy integrals
 \bea
 \left( P_{\Gamma}^{\pm} f \right) (\tau) = \pm \frac{1}{2 \pi i} \int_{\Gamma} \, \frac{f(z)}{z - \tau} \, dz  \;\;\;,\;\;\; \tau \in \mathbb{D}^{\pm}_{\Gamma} \;,
 \label{Ppm}
 \eea
where $ \mathbb{D}^+_{\Gamma}$ and  $\mathbb{D}^-_{\Gamma}$ denote the interior and the exterior regions of $\Gamma$, respectively.

$ P_{\Gamma}^+ f $ and $ P_{\Gamma}^- f$, defined by \eqref{Ppm}, are bounded analytic functions in $ \mathbb{D}^+_{\Gamma}$ and  
$\mathbb{D}^-_{\Gamma} \cup \{ \infty \}$,
respectively, with
\bea
\left( P_{\Gamma}^- f \right) (\infty) = 0 \;.
\eea
Their non-tangential boundary value limits on $\Gamma$ are uniquely defined a.e. \cite{MP}. It is usual to identify the functions defined on
$ \mathbb{D}^+_{\Gamma}$ and  $\mathbb{D}^-_{\Gamma}$ by \eqref{Ppm} with their respective boundary value function, which are also H\"older
continuous functions with exponent $\mu$.

Any element of $C^{\mu} (\Gamma)$ can be uniquely decomposed as $f = P_{\Gamma}^+ f + P_{\Gamma}^- f $.
If the winding number, or index, of $f \in C^{\mu} (\Gamma)$ is zero, then $\log f \in  C^{\mu} (\Gamma)$ and we can decompose
\bea
\log f = P_{\Gamma}^+ \log f + P_{\Gamma}^- \log f \;,
\eea
so that $f$ also admits a multiplicative decomposition which is a canonical Wiener-Hopf factorisation with factors in $C^{\mu} (\Gamma)$,
\bea
f = {\tilde f}_- {\tilde f}_+ \;\;\;,\;\;\; {\tilde f}_+^{\pm1} \in P_{\Gamma}^+  C^{\mu} (\Gamma) \;\;\;,\;\;\; {\tilde f}_-^{\pm1} \in P_{\Gamma}^-  C^{\mu} (\Gamma) \oplus \mathbb{C} \;,
\label{ffpfm}
\eea
where 
\bea
{\tilde f}_+ = e^{ P_{\Gamma}^+ \log f } \;\;\;,\;\;\; {\tilde f}_- = e^{ P_{\Gamma}^- \log f } \;\;\;,\;\;\; {\tilde f}_- (\infty) = 1 \;.
\label{fefefe}
\eea
The factors in \eqref{ffpfm} can be modified by a multiplicative constant 
($ {\tilde f}_+^{\pm} (0)$), and thus we can normalize ${\tilde f}_+$ to the identity at $0$, i.e. we can write
\bea
f = {f}_- {f}_+ \;\;\; \text{with} \;\;\; f_+ (0) = 1\;.
\label{fffn0}
\eea

It can be shown that a scalar function $f \in C^{\mu} (\Gamma)$ admits a canonical Wiener-Hopf factorisation if and only if its index with respect
to the origin is $0$, i.e. $\log f \in C^{\mu} (\Gamma)$ (see Lemma 2.1 in \cite{CG}).

Theorem \ref{theojoh} means that every $f_{\rho,v} (\tau)$ satisfying its assumptions is such that 
$\log f_{\rho,v} (\tau) \in C^{\mu} (\Gamma)$ and $f_{\rho,v} (\tau)$ admits a canonical Wiener-Hopf factorisation in the algebra $C^{\mu} (\Gamma)$ of the form \eqref{fffn0}.

For any diagonal matrix in $\left(  C^{\mu} (\Gamma) \right)^{n \times n}$,
\bea
{\cal M}(\tau) = {\rm diag} (m_j (\tau))_{j=1, \dots, n} \;,
\eea
where $m_j$ admits a canonical Wiener-Hopf factorisation in the algebra $C^{\mu} (\Gamma)$,  $m_j = m_j^- m_j^+$, we thus have the (normalized) 
canonical Wiener-Hopf factorisation
\bea
{\cal M} = {\cal M}_- \,  X  \;\;\, \textrm{on} \;\; \Gamma
\label{MMmX} 
\eea
with 
\bea
{\cal M}_- (\tau) &=& {\rm diag} (m^-_j (\tau) m^+_j (0))_{j=1, \dots, n} \;, \nonumber\\
X(\tau) &=& {\rm diag} (m^+_j (\tau) (m^+_j (0))^{-1} )_{j=1, \dots, n} \;,
\eea
where $X(0) = \mathbb{I}_{n \times n} $. 
We have thus proved the following:

{\theorem \label{theohc}
Let ${\cal M} (\omega)$ be a diagonal matrix function of the complex variable $\omega$ and let ${\cal M}_{\rho,v} (\tau)$ be its
composition with \eqref{spec}, such that it is H\"older continuous on a contour $\Gamma$.
Then ${\cal M}_{\rho,v} (\tau)$  has a canonical Wiener-Hopf factorisation in the algebra  $\left(  C^{\mu} (\Gamma) \right)^{n \times n}$ and
\bea
M(\rho, v) = e^{(P_{\Gamma}^+ \log {\cal M}_{\rho,v} (\tau) )_{(0)} }
\eea
is a solution to the field equation \eqref{fi2d}.\\
}

With the assumptions of Theorem \ref{theohc}, $X$ in \eqref{MMmX} is given by 
\bea
X(\tau, \rho, v)  =  e^{P_{\Gamma}^+ \log {\cal M}_{\rho,v} (\tau)} \, e^{-  (P_{\Gamma}^+ \log {\cal M}_{\rho,v} (\tau))_{(0)}  }  \;.
\eea

\section{The $\tau$-invariance problem and solution generation by multiplication  \label{sec:nsfd}}

Theorem \ref{theoraccr} states that once a canonical Wiener-Hopf factorisation of a monodromy matrix is obtained, 
we get a matrix $M(\rho, v)$ which gives a solution to \eqref{fi2d}, by simply calculating the limit of the ``minus" factor when $\tau$ tends to $\infty$,
see \eqref{defM}.

It is now natural to ask if the same can be achieved through other types of factorisation of the monodromy matrix. Indeed, in Section 7 of \cite{Aniceto:2019rhg}, a case study showed
that it is also possible to obtain solutions to the field equations by means of meromorphic factorisations \cite{CLS} of a monodromy matrix and subsequently taking the limit $\tau \rightarrow \infty$ of one of the factors in the factorisation.
It was 
moreover shown that the solution thus obtained (a Kasner solution to the field equations of General Relativity in four dimensions) was different from the one obtained through a canonical Wiener-Hopf factorisation of the same monodromy matrix.

Our approach for finding solutions to the Lax pair which are not necessarily analytic will be based on a new equation, which is not a 
Riemann-Hilbert problem and instead represents what can be called the $\tau$-invariance of an expression involving $X(\tau, \rho, v)$ and $M(\rho, v)$. The proof 
of this result, which is presented in Theorem \ref{theo3.3}, relies on the action of the two projections that we define next.

{\definition
Let
\bea
P_{\pm} = \frac12 \left( I \pm \lambda \, p_F \, \star \right) \;,
\eea
where $p_F$ is a fixed point of the involution $i_{\lambda} (\tau) = - \lambda/\tau$.
}

{\proposition
$P_{\pm}$ are complementary projections, i.e.
\bea
P_{\pm}^2 = P_{\pm} \quad , \quad P_+ + P_- = I \quad, \quad P_+ P_- = P_- P_+ = 0 , 
\eea
and they satisfy
\bea
P_+ - P_- = \lambda \, p_F \, \star \quad , \quad \star P_{\pm} = \mp p_F \, P_{\pm} \quad , \quad  2 P_{\pm}  d \rho = 
d \rho \mp p_F dv  .
\label{PPrel}
\eea

}

We can now formulate the main result of this section.

 {
\theorem
\label{theo3.3}

Let $\Gamma$ be an admissible contour and let $X(\tau, \rho, v)$ and $M(\rho, v)$ be matrix functions such that

(i) $X^{\pm 1}$ are analytic with respect to $\tau$ in a neighbourhood $O$ of $\Gamma$ (as in Theorem \ref{theoraccr}), for all $(\rho, v)$,

(ii) $X^{\pm 1}, M$ are of class $C^2$ with respect to $(\rho, v)$.

Let moreover
\begin{eqnarray} 
G_{M,X} (\tau, \rho, v) &=&
 \tau \, d M +  \frac{1}{\rho}\left[(\lambda-\tau^2)d\rho + 2\lambda \tau dv\right] 
M \frac{\partial X}{\partial \tau} X^{-1} 
\nonumber\\	&&  
	+ \frac{\tau^2+\lambda}{\tau} M \left(\frac{\partial X}{\partial \rho} d\rho + \frac{\partial X}{\partial v} dv \right)X^{-1} \;,
	\label{GMX}
\end{eqnarray}
where we omitted the dependence of $X$ and $M$ on $(\tau, \rho, v)$ and $(\rho,v)$, respectively.

If, for all $(\rho,v)$, 
\bea
\frac{\partial}{\partial \tau} G_{M,X} (\tau, \rho, v) = 0 \quad \text{on} \quad \Gamma,
\label{tderGt}
\eea
then $M(\rho,v)$ is a solution to the field equation \eqref{fi2d} and $X(\tau, \rho, v)$, with $\tau = \varphi \in {\cal T}$, is a solution to the linear system \eqref{varXA}.

}

\begin{proof}

If \eqref{tderGt} holds, then 
\bea
G_{M,X} (\tau, \rho, v) = G_{M,X} (\pm p_F, \rho, v) \equiv G(\rho,v),
\eea
and 
\bea
 G(\rho, v) &=& G_{M,X} (p_F, \rho, v) =
p_F \, d M +  \frac{2 \lambda}{\rho}\left[d\rho + p_F dv\right] 
M \frac{\partial X}{\partial \tau}\vert_{\tau = p_F}  X^{-1} \vert_{\tau = p_F} \nonumber\\
 \Leftrightarrow G(\rho, v) & = & p_F \, d M +  \frac{4 \lambda}{\rho}  \left( P_- d\rho \right) M \frac{\partial X}{\partial \tau}\vert_{\tau = p_F}  X^{-1} \vert_{\tau = p_F}
\end{eqnarray}
By applying $P_+$ to both sides of the last identity we get
\bea
P_+ G(\rho, v) = p_F \, P_+  d M .
\label{ppG}
\eea
Analogously, we evaluate $G(\rho, v) = G_{M,X}(-p_F, \rho, v) $ and apply$P_-$ to both sides, thus obtaining
\bea
P_- G(\rho, v) = - p_F \, P_-  d M .
\label{pmG}
\eea
{From} \eqref{ppG} and \eqref{pmG} we infer
\bea
G(\rho, v) = p_F \, P_+  d M - p_F \, P_-  d M = p_F \, \lambda p_F \star d M = - \star dM , 
\eea
and hence
\bea
 - \star dM = G_{M,X} (\tau, \rho, v).
 \label{stMG}
\eea
Substituting $\tau = \varphi \in {\cal T}$ we have
\bea
&& - \star dM = \varphi \, d M 
+ \frac{\lambda + \varphi^2}{\varphi} \, M \, d \left[ X(\tau, \rho, v)\vert_{\tau = \varphi} \right] X^{-1} (\varphi, \rho, v) 
\nonumber\\
&& \Leftrightarrow - \varphi M^{-1} \star dM  = \varphi^2 M^{-1} d M + (\lambda + \varphi^2) d \left[ X(\tau, \rho, v)\vert_{\tau = \varphi} \right] X^{-1} (\varphi, \rho, v) \;.
\eea
Denoting $A = M^{-1} d M, {\tilde X} (\rho, v) = X (\varphi, \rho, v) $, we have that
\bea
- \varphi \star A  = \varphi^2 A + (\lambda + \varphi^2) \, d {\tilde X} {\tilde X}^{-1}, 
\eea
which is equivalent to
\bea
\varphi \left( d {\tilde X} + A {\tilde X} \right) = \star d {\tilde X} 
\eea
 by Proposition 4.1 in \cite{Aniceto:2019rhg}.

\end{proof}

It is natural to ask whether there are actually matrices $X$ and $M$ satisfying the assumptions of Theorem \ref{theo3.3} and such that 
\eqref{tderGt} holds, and how to obtain them. One possible method to obtain such solutions is through canonical Wiener-Hopf factorisation. We have the following.

{\proposition

If $X(\tau, \rho, v)$ and $M(\rho,v)$ are obtained from a monodromy matrix ${\cal M}_{\rho, v} (\tau)$ by canonical Wiener-Hopf factorisation with respect to 
$\Gamma$, as in \eqref{canfacfac}-\eqref{noI}, then $X$ and $M$ satisfy \eqref{tderGt}.

}

\begin{proof}

Indeed we have, in that case, (see equation (6.17) in \cite{Aniceto:2019rhg}) that
\bea
G_{M,X}^{\natural} \left( - \frac{\lambda}{\tau}, \rho, v \right) = G_{M,X} (\tau, \rho, v) \quad \text{on} \quad \Gamma,
\label{GGhash} 
\eea
where the left hand side is analytic and bounded in $\mathbb{D}_{\Gamma}^-$ and the right hand side is analytic and bounded in 
$\mathbb{D}_{\Gamma}^+$. Hence both sides must be equal to a constant. Note that  $G_{M,X} (\tau, \rho, v) $ has an apparent singularity at $\tau =0$,
but due to the normalization condition $X(0, \rho, v) = \mathbb{I}$ for all $\rho, v$, which implies that $\partial X/\partial \rho \vert_{\tau =0} = 
\partial X/\partial v \vert_{\tau =0} = 0$, this singularity is removable. Analogously, the apparent singularity of 
$G_{M,X} \left( - \frac{\lambda}{\tau}, \rho, v \right)$ at $\tau = \infty$ is removable.

\end{proof}

{\remark 
\label{remGK}

Note that a pair  $(M,X)$ that satisfies \eqref{tderGt} may not be associated with a matrix factorisation problem. However, suppose that 
\eqref{tderGt} is satisfied, in which case $G_{M,X} (\tau, \rho, v) = K(\rho, v)$,
where the matrix $K$ is independent of $\tau $, and assume moreover that $K = K^{\natural}$. Then we have
\bea
G_{M,X} (\tau, \rho, v) = G^{\natural}_{M,X} (\tau, \rho, v) = G^{\natural}_{M,X} \left(- \frac{\lambda}{\tau}, \rho, v\right) \quad \text{on} \quad \Gamma.
\eea
Note that we have  $K = K^{\natural}$ whenever $K$ is a diagonal matrix.
By the proof of Theorem 6.1 in \cite{Aniceto:2019rhg}, this is equivalent to 
\bea
d \left(  \left( X^{\natural}  \left(- \frac{\lambda}{\tau}, \rho, v \right) \, M(\rho, v) \, X (\tau, \rho, v) \right)\vert_{\tau=\varphi_{\omega}} \right)  = 0 \;,
\eea
and hence the product $\left( X^{\natural}  \left(- \frac{\lambda}{\tau}, \rho, v \right) \, M(\rho, v) \, X (\tau, \rho, v) \right)\vert_{\tau=\varphi} $
is a matrix ${\cal M}(\omega)$ that is independent
of $(\rho,v)$. This corresponds to obtaining a factorisation for a monodromy matrix $\cal M$ which is not, in general, a Wiener-Hopf factorisation, but
still provides a solution to the field equation \eqref{fi2d}. Indeed there exist solutions of the field equations which cannot be obtained from a canonical Wiener-Hopf factorisation of a monodromy matrix, such as the Kasner solution discussed in {\cite [Section 7]{Aniceto:2019rhg}}, but can be obtained from a factorisation of the type
mentioned above.\\

}

Motivated by this, 
we will now look for functions $R (\tau, \rho, v), N(\rho,v)$ which satisfy conditions $(i)-(ii)$ of Theorem
\ref{theo3.3}, with 
\bea
N(\rho, v) = N^{\natural} (\rho,v) ,
\eea
such that the product
\bea
R^{\natural} \left(\frac{- \lambda}{\tau}, \rho, v \right) \,N(\rho,v) R (\tau, \rho, v) \vert_{\tau = \varphi_{\omega} (\rho, v)}
\eea
is a $\natural$-invariant function that does not depend on $(\rho, v)$. We now proceed to characterize a wide class of such functions $R$ and $N$, 
such that all the assumptions of Theorem \ref{theo3.3} are satisfied and thus   \eqref{tderGt} holds.

For each $(\rho, v) \in S$, let $\omega_i \in \mathbb{C}$ with $\omega_i \neq v$ for all $(\rho, v) \in S$, and let
 $\tau_i(\rho,v)$ and ${\tilde \tau}_i (\rho,v) = - \frac{\lambda}{\tau_i(\rho,v)}$ be the two roots of 
\bea
\omega_i = v + \frac{\lambda}{2}  \,\rho \, \frac{\lambda - \tau^2}{\tau} .
\eea
For any admissible contour $\Gamma$ in the complex $\tau$-plane such that $\tau_i, {\tilde \tau}_i$ do not lie on the contour (note that 
 $\tau_i, {\tilde \tau}_i \neq \pm \sqrt{- \lambda}$, which are the two points that 
necessarily belong to any admissible contour), one of those points will 
be in the interior of $\Gamma$ and the other in the exterior (cf. \cite{Aniceto:2019rhg});
their relative position depends on the choice of $\Gamma$.
Now define
\bea
R^i (\tau, \rho, v) =  \frac{\tau_i}{ {\tilde \tau}_i} \,  \frac{\tau -  {\tilde \tau}_i}{\tau - \tau_i} .
\label{Ri}
\eea
Since $R^i$ is a scalar function, we have $\left( R^i\right)^{\natural} = R^i$.

{\proposition
 \label{propNi} 
Let $R^i (\tau, \rho, v)$ be any function defined as in \eqref{Ri} for some $\omega_i \in \mathbb{C}$ and let 
\bea
N_i (\rho,v) = \frac{{\tilde \tau}_i}{ {\tau}_i} \;, 
\label{alta1}
\eea
Then $R^i$ and $N_i$ satisfy the $\tau$-invariance condition \eqref{tderGt}.

}

\begin{proof}

Clearly, $R^i$ and $N_i$ satisfy the assumptions $(i)-(ii)$ of Theorem \ref{theo3.3}, taking into account that
\bea
\left(  \frac{\tau -  {\tilde \tau}_i}{\tau - \tau_i} \right)^{-1} 
= \frac{\tau_i}{ {\tilde \tau}_i} \; \; \frac{ - \frac{\lambda}{\tau} - {\tilde \tau}_i}{- \frac{\lambda}{\tau} - { \tau}_i}
\;.
\eea
Moreover, we have that 
\bea
(R^i)^{\natural} (\frac{- \lambda}{\tau}, \rho, v) \,N_i(\rho,v) R^i (\tau, \rho, v) = 1 ,
\label{RNRid}
\eea
so $N_i = (N_i)^{\natural}$ and, substituting $\tau = \varphi \in {\cal T}$, it follows that 
\bea
d \left(  \left( (R^i)^{\natural} N_i R^i \right)\vert_{\tau = \varphi} \right) = 0 ,
\eea
where $d$ denotes the differential with respect to $(\rho, v)$. Now, following the first part of the proof of Theorem 6.1 in \cite{Aniceto:2019rhg}, it is easy to see that
this implies that, in a neighbourhood of $\Gamma$, 
\begin{eqnarray} 
	&&  - \frac{\lambda}{\tau^2\rho}[(\lambda-\tau^2)d\rho+2\lambda \tau dv]  \left(R^{i \natural} \right)^{-1} \left(-\frac{\lambda}{\tau},\rho,v \right)
	 \left.\frac{\partial R^{i \natural} (\tau',\rho,v)}{\partial \tau'}  \right\vert_{\tau'=-\frac{\lambda}{\tau}} N_i(\rho,v)
		 \nonumber\\
	 && -\frac{\tau^2+\lambda}{\tau}   \left(R^{i \natural} \right)^{-1}\left(-\frac{\lambda}{\tau},\rho,v \right)\left[\frac{\partial R^{i \natural}}{\partial \rho}\left(-\frac{\lambda}{\tau},\rho,v \right)d\rho + \frac{\partial R^{i \natural}}{\partial v}\left(-\frac{\lambda}{\tau},\rho,v \right) dv \right] N_i(\rho,v) \nonumber\\
	&&  -\frac{\lambda}{\tau} d N_i (\rho,v) \nonumber\\
&& =	 \tau \, dN_i (\rho,v) +  \frac{1}{\rho}\left[(\lambda-\tau^2)d\rho + 2\lambda \tau dv\right] 
N_i(\rho,v) \frac{\partial R^i}{\partial \tau}(\tau,\rho,v) \, (R^i)^{-1}(\tau,\rho,v) \nonumber\\
	&&  + \frac{\tau^2+\lambda}{\tau} N_i(\rho,v) \left(\frac{\partial R^i}{\partial \rho}(\tau,\rho,v)d\rho 
	+ \frac{\partial R^i}{\partial v}(\tau,\rho,v)dv \right)(R^i)^{-1}(\tau,\rho,v) \;.
		\label{2rhpR}
\end{eqnarray}
We have on the right hand side,
\bea
\label{longexp}
&&  \tau \, dN_i +  N_i \frac{1}{\rho}  \left[ \left[(\lambda-\tau^2)d\rho + 2\lambda \tau dv\right] 
 \frac{\partial R^i}{\partial \tau}  
  + \frac{\tau^2+\lambda}{\tau}  \, \rho \,  \left(\frac{\partial R^i}{\partial \rho} d\rho + \frac{\partial R^i}{\partial v} dv \right) \right](R^i)^{-1} 
  \nonumber\\
  && = \tau \, dN_i +  N_i   \frac{1}{\rho}  
  \left[ ( \lambda - \tau^2)  \frac{\partial R^i}{\partial \tau}   +  \frac{\tau^2+\lambda}{\tau} \, \rho \, 
   \frac{\partial R^i}{\partial \rho}   \right]  (R^i)^{-1} d \rho   \nonumber\\
   &&  \quad  +  N_i   \frac{1}{\rho}   \left[ 2 \lambda \tau \,  \frac{\partial R^i}{\partial \tau }
+ \frac{\tau^2+\lambda}{\tau}  \, \rho \,  \frac{\partial R^i}{\partial v}   \right] (R^i)^{-1}  dv . 
   \eea
The first term on the right hand side of \eqref{longexp} is analytic and bounded in $\mathbb{D}^+_{\Gamma} \cup O$,
where $O$ is a neighbourhood of $\Gamma$. The second and third terms have an apparent pole at $\tau =0$, which is 
cancelled by the fact that $R^i (0, \rho, v) = 1$ implies that 
$ \frac{\partial R^i}{\partial \rho} $ and  $\frac{\partial R^i}{\partial v} $ vanish at $\tau =0$. Since $(R^i)^{\pm 1}$ are analytic in 
$\left(\mathbb{D}^+_{\Gamma} \cup O \right) \backslash \{ \tau_i, {\tilde \tau}_i \}$, what is left is to analyse the behaviour of those two terms in a neighbourhood of 
 $\tau = \tau_i, {\tilde \tau}_i $, to show that 
 \bea
   \left[ ( \lambda - \tau^2)  \frac{\partial R^i}{\partial \tau} \,  +  \frac{\tau^2+\lambda}{\tau} \, \rho \, 
   \frac{\partial R^i}{\partial \rho} \right] (R^i)^{-1}
   \label{Rirel} 
   \eea
 and 
 \bea
 \left[ 2  \lambda \tau  \frac{\partial R^i}{\partial \tau} \,  +  \frac{\tau^2+\lambda}{\tau} \, \rho \, 
   \frac{\partial R^i}{\partial v}   \right] (R^i)^{-1}
    \label{Rirel2} 
\eea
have only apparent singularities at those points.  From  \eqref{Ri} we have that
 \bea
&& \frac{\partial R^i}{\partial \tau} = \frac{\tau_i}{ {\tilde \tau}_i }\,  \frac{{\tilde \tau}_i -  { \tau}_i}{(\tau - {\tau}_i)^2} \;\;\;,\;\;\;
\nonumber\\
&& 
\frac{\partial R^i}{\partial \rho} =  \frac{\partial R^i}{\partial \tau_i}   \frac{\partial \tau_i}{\partial \rho} = - \lambda \left( \frac{\tau_i \tau + \lambda}{\tau - \tau_i}
 + \tau_i \frac{\lambda + \tau^2}{(\tau - \tau_i)^2} \right) \frac{\tau_i}{\rho} \,  \frac{ \lambda - \tau_i^2}{\lambda + \tau_i^2} ,
\eea
and taking  into account that $\lambda^2=1$ and ${\tilde \tau}_i = - \lambda /\tau_i$, one finds by direct calculation that  \eqref{Rirel}
has removable singularities at 
$\tau = \tau_i, {\tilde \tau}_i $ since the numerator vanishes at those points.
Likewise, \eqref{Rirel2} also has removable singularities at  $\tau = \tau_i, {\tilde \tau}_i $.  

We therefore conclude that the right hand side of \eqref{2rhpR} represents a function 
which is analytic and bounded in $\mathbb{D}^+_{\Gamma}$ and, analogously, the left hand side of \eqref{2rhpR} 
represents a function which is analytic and bounded in $\mathbb{D}^-_{\Gamma}$, so both sides must be equal to a constant (with respect to $\tau$).

\end{proof}

The following is a straightforward consequence of the above.

{\corollary
\label{corR}
For any set of points $\omega_i, i = 1, \dots, n,$ 
such that  $\omega_i \neq v $   
for all $(\rho, v) \in S$, let $R(\tau, \rho, v) = 
{\rm diag} (R_i (\tau, \rho, v))_{i=1,  \dots, n}$, where $R_i$ is defined by \eqref{Ri}, and let $N = {\rm diag} (N_i)_{i=1,  \dots, n}$ with $N_i$ given
by \eqref{alta1}. Then $R(\tau, \rho, v) $ and $N(\rho, v)$ satisfy \eqref{tderGt}, so 
$R(\tau, \rho, v)\vert_{\tau = \varphi_{\omega}}$ is a solution to the linear system \eqref{varXA}  for $A = N^{-1} d N$, and $N(\rho, v)$
is a solution to the field equation \eqref{fi2d}.\\}

It is easy to see that the following also holds.

{\proposition
\label{propRal}

The conclusions of Corollary \ref{corR}  still hold if we replace $R^i$ by any product of functions of the form \eqref{Ri}, or
by $(R^i)^{\alpha}$ with $\alpha \in \mathbb{R}$, with the corresponding change in \eqref{alta1}.
\\}

Note that we have
 \bea
 \mathbb{I}_{n\times n} = R^{\natural} \left(\frac{- \lambda}{\tau}, \rho, v \right) \,N(\rho,v) R (\tau, \rho, v) \;,
 \label{IRNR2}
 \eea
in agreement with Remark \ref{remGK}.

The next theorem shows that we can use the solutions $N(\rho,v)$ to generate, from a given solution $M(\rho,v)$ of
 \eqref{fi2d}, a new family of solutions obtained by multiplication, provided that certain commutation relations involving \underline{both} the solutions
 to  \eqref{fi2d} and the corresponding solutions to the linear system \eqref{varXA} hold.

{\theorem
\label{theoMN}

Let $\Gamma$ be an admissible contour and let $R(\tau, \rho, v)$ and $N(\rho,v)$ be $n \times n$ matrix functions such that

(i) $R^{\pm 1}$ are analytic with respect to $\tau$ in a neighbourhood $O$ of $\Gamma$, for all $(\rho, v)$;

(ii) $R^{\pm 1}$ are of class $C^2$ with respect to $(\rho, v)$;

(iii) $N = N^{\natural}$ and $N^{\pm 1}$ are of class $C^2$ with respect to $(\rho, v)$;

(iv) $G_{N,R}$ (defined by \eqref{GMX}, with $M$ and $X$ replaced by $N$ and $R$, respectively),
does not depend on $\tau$ for any $\tau \in \Gamma$, for all $(\rho, v) \in S$. 

Let moreover $X(\tau, \rho, v)$ and $M(\rho, v)$ be $n \times n$ matrix functions satisfying the same conditions as $R$ and $N$, respectively.
Then, if $R$ commutes with $X$ and its derivatives with respect to $\tau, \rho, v$; $M$ commutes with $R$ and its derivatives with respect to 
$\tau, \rho, v$; $N$ commutes with $X$ and its derivatives with respect to $\tau, \rho, v$, we have that $(R X)\vert_{\tau = \varphi_{\omega}}$ is a 
solution to the linear system \eqref{varXA} for $A = (M N )^{-1} d (M N)$ and $(M N) (\rho, v)$ is a solution to  \eqref{fi2d}.

}

\begin{proof} 

By straightforward calculation one sees that if the assumptions hold, in particular the commutation relations, then
(omitting the dependence on the variables $\tau, \rho, v$ for notational simplicity)
\begin{eqnarray}
\label{p2RHMN}
&& G_{MN, RX} (\tau, \rho, v) \nonumber\\
&& =	 \tau \, d(M N) (\rho,v) +  \frac{1}{\rho}\left[(\lambda-\tau^2)d\rho + 2\lambda \tau dv\right] 
(M N)(\rho,v) \frac{\partial (R X)}{\partial \tau}(\tau,\rho,v) ( R X) ^{-1}(\tau,\rho,v) \nonumber\\
	&&  + \frac{\tau^2+\lambda}{\tau} ( M N)(\rho,v) \left(\frac{\partial (R X)}{\partial \rho}(\tau,\rho,v)d\rho + \frac{\partial (R X) }{\partial v}(\tau,\rho,v)dv \right) (R X)^{-1}(\tau,\rho,v) \nonumber\\
&& = 
 \tau \, (dM) N  +  \frac{1}{\rho}\left[(\lambda-\tau^2)d\rho + 2\lambda \tau dv\right] 
M N \frac{\partial X}{\partial \tau} R   R^{-1} X^{-1}  \nonumber\\
&& + \frac{\tau^2+\lambda}{\tau}  M N \left(\frac{\partial X}{\partial \rho} d\rho + \frac{\partial X }{\partial v} dv \right) R R^{-1} X^{-1}
	\nonumber\\
&& + 
 \tau \, M dN +  \frac{1}{\rho}\left[(\lambda-\tau^2)d\rho + 2\lambda \tau dv\right] 
M N X  \frac{\partial R }{\partial \tau} R^{-1} X^{-1} \nonumber\\
&&  + \frac{\tau^2+\lambda}{\tau} M N X \left(\frac{\partial R }{\partial \rho} d\rho + \frac{\partial R  }{\partial v} dv \right) R^{-1} X^{-1} \nonumber\\
&& = \left[ 
 \tau \, dM  +  \frac{1}{\rho}\left[(\lambda-\tau^2)d\rho + 2\lambda \tau dv\right] 
M  \frac{\partial X}{\partial \tau} X^{-1} 
	  + \frac{\tau^2+\lambda}{\tau}  M  \left(\frac{\partial X}{\partial \rho} d\rho + \frac{\partial X }{\partial v} dv \right)  X^{-1}  \right] N
	\nonumber\\
&& + M \left[
 \tau \, dN +  \frac{1}{\rho}\left[(\lambda-\tau^2)d\rho + 2\lambda \tau dv\right] 
 N   \frac{\partial R }{\partial \tau} R^{-1} 
	  + \frac{\tau^2+\lambda}{\tau}  N  \left(\frac{\partial R }{\partial \rho} d\rho + \frac{\partial R  }{\partial v} dv \right) R^{-1} 
	  \right] \nonumber\\
&& =  G_{M,X} (\tau, \rho, v) \, N(\rho, v) + M(\rho,v) \, G_{N,R}(\tau, \rho,v) ,
\end{eqnarray}
where we have made use of the definition given in \eqref{GMX}. Since, by assumption,  $G_{M,X}$ and $G_{N,R}$ 
satisfy the $\tau$-invariance condition \eqref{tderGt}, also $G_{MN, RX} $ satisfies the $\tau$-invariance condition.
The result then follows from Theorem \ref{theo3.3}.

\end{proof}

{\remark \label{remMXRN}
The assumptions of Theorem \ref{theoMN}, in particular the commutation relations, are naturally satisfied if the matrices involved are diagonal.
\\
 }


Before proceeding, let us consider a concrete example of the solution generating method presented in Theorem \ref{theoMN}.

\subsection{From the interior Schwarzschild solution to a cosmological Kasner solution}

As an application of the solution generating method presented above,  we show how to relate the interior region of the Schwarzschild solution, obtained
through the canonical Wiener-Hopf factorisation of a monodromy matrix with respect to $\Gamma$, 
to a particular cosmological Kasner solution.

We consider the Schwarzschild monodromy matrix
\bea
{\cal M} (\omega) = 
\begin{pmatrix}
\lambda \, \frac{\omega - {m} }{\omega + {m}} & 0 \\
0 &  \lambda \, \frac{\omega +{m} }{\omega - {m}}
\end{pmatrix} \;.
\label{monschwarz}
\eea
The matrix ${\cal M}_{\rho, v} (\tau)$ which results by composing \eqref{monschwarz} with
\eqref{spec} reads \cite{Aniceto:2019rhg}
\bea
{\cal M}_{\rho, v} (\tau) = 
\begin{pmatrix}
\lambda \, \frac{(\tau -  \varphi_m)(\tau + \lambda/ \varphi_m) }{(\tau -  \varphi_{-m})(\tau + \lambda/ \varphi_{-m})} & 0 \\
0 &  \lambda \,  \frac{(\tau -  \varphi_{-m})(\tau + \lambda/ \varphi_{-m}) }{(\tau -  \varphi_{m})(\tau + \lambda/ \varphi_{m})}
\end{pmatrix} \;,
\label{schwmon}
\eea
where $\varphi_m (\rho, v) $ and $ \varphi_{-m} (\rho, v)$ are given by
\bea
\varphi_{\alpha} (\rho, v) = - \lambda \, \frac{(\alpha - v) + \sqrt{ (\alpha - v)^2 + \lambda \rho^2}}{\rho} \;\;\;,\;\;\; \alpha \in \mathbb{C} \;.
\label{defvarp}
\eea
The canonical Wiener-Hopf factorisation of ${\cal M}_{\rho, v} (\tau)$, for both $\lambda = 1$ and $\lambda = -1$, was discussed in detail in \cite{Aniceto:2019rhg}.
Depending on the choice of the contour $\Gamma$, the canonical factor $M_c (\rho, v)$ either describes the exterior region ($\lambda = 1$) or the interior
region ($\lambda = -1$) of the Schwarzschild solution, or yet another solution, see {\cite [Section 8.1]{Aniceto:2019rhg}}.
In all these cases, the resulting canonical factor
\bea
M_c (\rho, v) = 
 \begin{pmatrix}
\Delta_c (\rho, v)
 & 0
 \\
 0 &\Delta^{-1}_c (\rho, v)
  \end{pmatrix} 
  \eea
is determined in terms of ratios or products of $\varphi_m$ and $\varphi_{-m}$, i.e. $\Delta_c = \lambda \,  (\varphi_m)^a \,  (\varphi_{-m})^b$ with $a, b = \pm 1$ 
 \cite{Aniceto:2019rhg}.

In the following, we take $\lambda= -1$ and, following \cite{Aniceto:2019rhg},
consider the solution that describes the interior of the Schwarzschild
solution in region $I$ of the Weyl upper-half plane. This is the region in the $(\rho, v)$ plane (with $\rho > 0$)
 delimited by the following three lines: a) $ \rho = 0, -m < v < m$, b) $\rho = v +m$ and c)
$\rho = m - v$. This solution arises 
by canonically factorising \eqref{schwmon} with respect to a contour $\Gamma$ in the complex $\tau$-plane that passes through the fixed points
$\tau = \pm 1$ of the involution $\tau \mapsto 1/\tau$. We consider a contour $\Gamma$ such that 
\bea
{\tilde \tau}_1 = \frac{m - v + \sqrt{ (m-v)^2 - \rho^2}}{\rho} \;\;\;,\;\;\; {\tilde \tau}_2 = \frac{- m - v + \sqrt{ (m+v)^2 - \rho^2}}{\rho} 
\eea
lie inside the contour $\Gamma$, and $\tau_1 = 1/ {\tilde \tau}_1$ and $\tau_2 = 1/ {\tilde \tau}_2$ lie outside of $\Gamma$. 
Note that $\tau_1>0, {\tilde \tau}_1 > 0$, while $\tau_2 < 0, {\tilde \tau}_2 <0$.
Then, the canonical Wiener-Hopf factorisation
of ${\cal M}_{\rho, v} (\tau)$ with respect to this contour yields \cite{Aniceto:2019rhg}
\bea
X_c = \begin{pmatrix}
\frac{{\tilde \tau}_1}{{\tilde \tau}_2} \, \left(\frac{ \tau - \tau_1}{\tau - \tau_2}\right)
 & 0
 \\
 0 & \frac{{\tilde \tau}_2}{{\tilde \tau}_1} \, \left(\frac{ \tau - \tau_2}{\tau - \tau_1}\right)
 \end{pmatrix} \;\;\;,\;\;\;
 M_c = \begin{pmatrix}
- \frac{{\tilde \tau}_2}{{\tilde \tau}_1}
 & 0
 \\
 0 &- \frac{{\tilde \tau}_1}{{\tilde \tau}_2} 
  \end{pmatrix} \;.
  \label{intschw}
 \eea
  Now we choose matrices $R (\tau, \rho, v)$ and $N(\rho,v)$ according to Proposition \ref{propRal}, given by
 \bea
R(\tau, \rho, v) &=& \begin{pmatrix}
 \left(   \frac{ \tau_1^2}{\tau_2^2}   \, \left( \frac{\tau - {\tilde \tau}_1 }{\tau - \tau_1 } \right) 
\left( \frac{\tau - \tau_2 }{\tau - {\tilde \tau_2 }} \right)  \right)^{1/2} & 0 \\
0 &
\left(   \frac{ \tau_2^2}{\tau_1^2}   \, \left( \frac{\tau - {\tilde \tau}_2 }{\tau - \tau_2 } \right) 
\left( \frac{\tau - \tau_1 }{\tau - {\tilde \tau_1 }}  \right) \right)^{1/2}
\end{pmatrix}
\;, \nonumber\\
 N(\rho, v) &=&  \begin{pmatrix}
 \sqrt{\frac{ {\tilde \tau}_1^2}{{\tilde \tau}_2^2}} & 0 \\
 0 &  \sqrt{\frac{ {\tilde \tau}_2^2}{{\tilde \tau}_1^2}} 
 \end{pmatrix}
 \;.
 \label{Rdefschw}
  \eea
 Multiplying as in Theorem \ref{theoMN}, we obtain
     \bea
M_m (\rho, v) = 
 M_c (\rho,v) \, N(\rho,v)
= \begin{pmatrix}
1
 & 0
 \\
 0 &1
  \end{pmatrix}  = \begin{pmatrix}
\Delta_m
 & 0
 \\
 0 &\Delta_m^{-1} 
  \end{pmatrix} \;,
  \eea
where we used $\sqrt{ {\tilde \tau}_1^2 / {\tilde \tau}_2^2} = - {\tilde \tau}_1/ {\tilde \tau}_2$.
 The associated matrix one-from $A_m = M^{-1}_m d M_m$ vanishes, and the 
factor $\psi$, obtained by integrating \eqref{eq_psi}, is constant. We take it to be zero. The associated space-time metric \eqref{4dWLP}
takes the form 
 \bea
ds_4^2 = 
 -d \rho^2 + dv^2 +  \rho^2 d\phi^2 + dy^2 \;,
 \label{kasn010}
\eea
which describes a cosmological Kasner solution with exponents $(p_1, p_2, p_3) = (0, 1, 0)$.
Note that the Killing horizon of \eqref{intschw}, which is located at $ \rho = 0, -m < v < m$, gets mapped to the cosmological singularity $\rho = 0$ of 
\eqref{kasn010}.

In the next sections we apply the previous results to several diagonal solutions $M(\rho, v)$ of \eqref{fi2d}.

\section{Cosmological Kasner solutions and Kac-Moody currents \label{sec:cosmk}}

Let us consider here the simplest kind of diagonal monodromy matrices, those 
associated to ${\cal M} (\omega)$ with monomial diagonal elements and determinant $1$, of the form
\bea
\mathcal{M} (\omega) = \begin{bmatrix}
	   (\omega - a)^{N}
		\; \; & \;\;	0 \\
		0\;  \; & \; \;	   (\omega - a)^{-N}
\end{bmatrix} \;\;\;, \;\;\; N  \in \mathbb{N} \;\;\;,\;\;\; a\in \mathbb{R} \;.
\label{monN}
\eea
We consider the Einstein field equations in vacuum in four space-time dimensions, in which case $\natural$ is simply matrix transposition, and 
we take $\lambda = -1$. In \cite{Aniceto:2019rhg} the case of a monodromy matrix of this type, namely ${\rm diag} \left( \omega^4, \omega^{-4} \right)$,
was studied, showing that one could obtain, by meromorphic factorisation, a solution to \eqref{fi2d} depending only on $\rho$, which was a particular
Kasner solution. We address here the question of whether this can be extended to the general class of monodromy matrices of the form \eqref{monN}.

\begin{figure}
\hskip -5mm
	\includegraphics[align=c, scale=0.52]{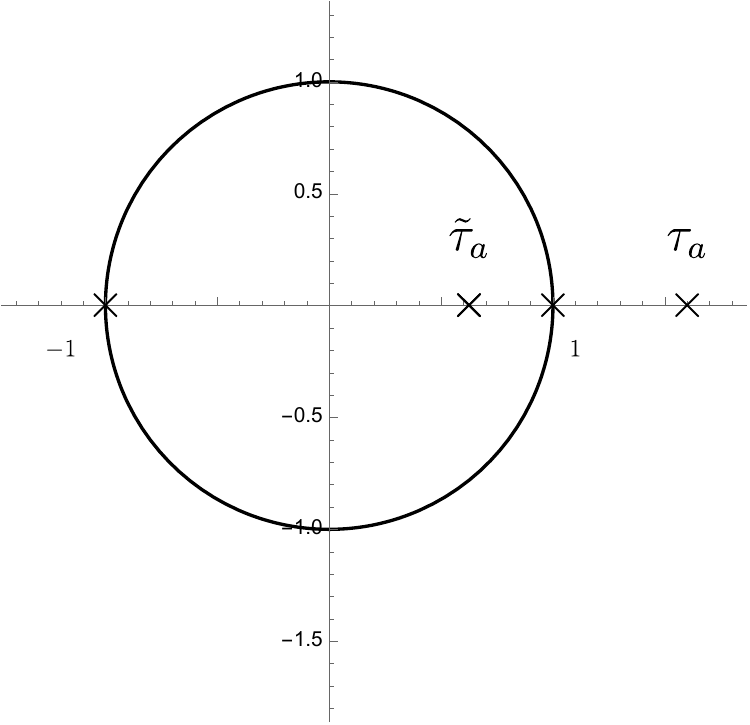} \hskip 12mm
	\includegraphics[align=c, scale=0.52]{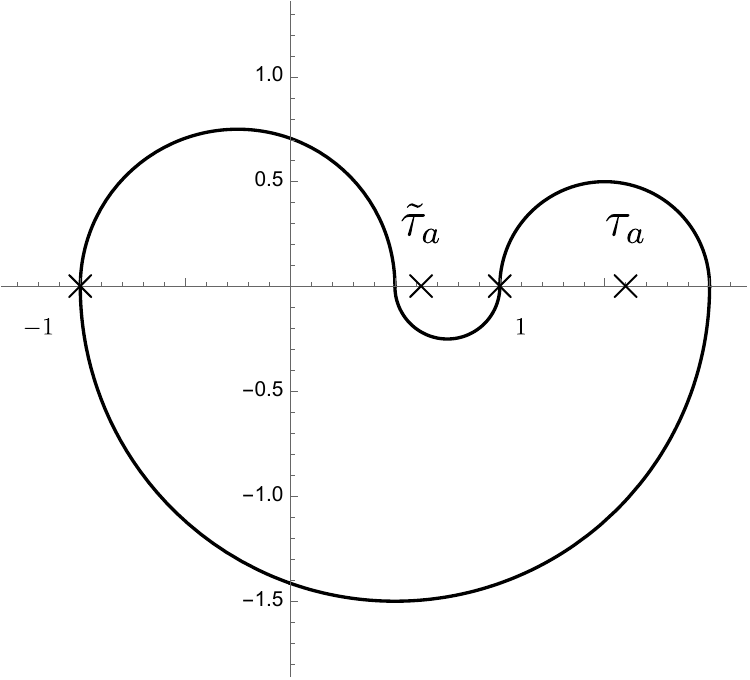}
			\caption{Examples of a contour
						for $v=0, \rho=1, a = \frac{3.56}{3.2}$ ($\tau_a = 1.6, {\tilde \tau}_a = 0.625$),
			with either $\tau_a$ or  ${\tilde \tau}_a$ in its interior.
		\label{cont}}
\end{figure}

Note that, composing $\mathcal{M} (\omega)$ with \eqref{spec}, we obtain a  rational matrix ${\cal M}_{\rho, v} (\tau)$, for which 
a canonical Wiener-Hopf factorisation with respect to any curve $\Gamma$ not containing its zeros and poles can be immediately obtained.
${\cal M}_{\rho, v} (\tau)$ takes the form
\bea
{\cal M}_{\rho, v} (\tau) &=& 
\begin{bmatrix}
	\left( (v - a + \frac12 \rho \frac{(1 + \tau^2)}{\tau} \right)^N & 0 \\
	0 & \left( (v - a + \frac12 \rho \frac{(1 + \tau^2)}{\tau} \right)^{-N}
\end{bmatrix} \nonumber\\
&=&
\begin{bmatrix}
	(\frac12 \rho)^N \left(  \frac{(\tau - \tau_a)(\tau - {\tilde \tau}_a) }{\tau}  \right)^N & 0 \\
	0 & 2^N \rho^{-N}  \left(  \frac{(\tau - \tau_a)(\tau - {\tilde \tau}_a) }{\tau}  \right)^{-N} 
\end{bmatrix} ,
\label{monkas}
\eea
with 
\bea
\tau_a = \frac{-  (v - a) +  \sqrt{(v-a)^2 - \rho^2}}{\rho} \;\;\;,\;\;\; {\tilde \tau}_a = \frac{-  (v - a) -  \sqrt{(v-a)^2 - \rho^2}}{\rho} .
\eea
For any admissible contour $\Gamma$ in the complex $\tau$-plane not containing $\tau_a$ and ${\tilde \tau}_a$,
one of these two points will be in the interior region of $\Gamma$ and the other will be in the exterior region, see Proposition 3.4 in \cite{Aniceto:2019rhg}. On the other hand, it is possible
to define simple closed contours $\Gamma$ for which either point is in the interior  (see Figure \ref{cont}).
Note that $\tau_a$ and ${\tilde \tau}_a$ have real parts with the same sign, and $\tau_a, {\tilde \tau}_a \neq \pm 1$.

Let us consider a contour $\Gamma_{\tau_a}$, with $\tau_a$ in its interior region. Then a canonical Wiener-Hopf factorisation of ${\cal M}_{\rho, v} (\tau) $ is
given by
\bea
\label{canKas}
&&{\cal M}_{\rho, v} (\tau) =  X^{\natural} ( - \frac{\lambda}{\tau}, \rho, v) \, M(\rho, v) \, X(\tau, \rho, v)  \\
&&= \begin{bmatrix}
	\left(  \frac{(\tau - \tau_a) }{\tau}  \right)^N & 0 \\
	0 &  \left(  \frac{(\tau - \tau_a) }{\tau}  \right)^{-N} 
	\end{bmatrix} 
	\begin{bmatrix}
	 \left(- \frac12 \rho {\tilde \tau}_a \right)^N  & 0 \\
	0 & \left(- \frac12 \rho {\tilde \tau}_a \right)^{-N} 
	\end{bmatrix}
	\begin{bmatrix}
	\left(  \frac{\tau - {\tilde \tau}_a }{- {\tilde \tau}_a}  \right)^N & 0 \\
	0 & \left(  \frac{\tau - {\tilde \tau}_a }{- {\tilde \tau}_a}  \right)^{-N} 
	\end{bmatrix},
	\nonumber
\eea
leading to the solution 
\bea
M_a (\rho, v) =  {\rm diag} \left( \left(- \frac12 \rho {\tilde \tau}_a \right)^N,   \left(- \frac12 \rho {\tilde \tau}_a \right)^{-N} \right).
\label{Mac}
\eea
Taking into account that 
$
 {\tilde \tau}_a^2 =\frac{{\tilde \tau}_a}{\tau_a}
$,
we conclude from Remark \ref{remMXRN} that by multiplying \eqref{Mac} by (cf. \eqref{alta1} and Proposition \ref{propRal})
\bea
{\rm diag}\left(  \left( \frac{\tau_a}{{\tilde \tau}_a} \right)^n,  \left( \frac{{\tilde \tau}_a}{\tau_a} \right)^{n} \right) \;\;\;,\;\;\;  n = \frac{N}{2},
\eea
we get a solution depending only on $\rho$, 
\bea
M_m(\rho,v) =   {\rm diag} \left( \left( \frac12 \rho \right)^{2n}, \left(\frac12 \rho \right)^{-2n} \right) ,
\label{Mmk}
\eea
which is independent of the choice of $a$.

Taking the four-dimensional line element  to be of the form \eqref{4dWLP} with 
\bea
ds_2^2 =  - d \rho^2 + dv^2 ,
\label{kmrv}
\eea
using 
$\Delta (\rho) =  (\frac12 \rho)^{2n} $ and
$
e^{\psi (\rho)} = c \, \rho^{2 n^2} $ $(c \in \mathbb{R}$), and defining the rational numbers
\bea
p_1 = \frac{n (n-1) }{n^2 - n + 1} \;\;\;,\;\;\; p_2 = \frac{1-n}{n^2 -n+1} \;\;\;,\;\;\; p_3 = \frac{n}{n^2-n +1} 
\label{p1p2p3}
\eea
(which satisfy $p_1 + p_2 + p_3 = 1, \; p_1^2 + p_2^2 + p_3^2 = 1$ with $0 \leq p_1< 1$),
we obtain, after the change of coordinates 
\bea
 \rho = t^{1 -p_1} \;\;\;,\;\;\; v = (1-p_1) \, x_1 \;\;\;,\;\;\; t >0 , 
\eea
the four-dimensional line element
\bea
ds_4^2 =   - dt^2 + t^{2 p_1}  \, d x_1^2 + t^{2 p_2}  \, d x_2^2 +  t^{2 p_3}  \, d x_3^2  \;,
\label{solk}
\eea
where 
we chose the integration constant $c$ to equal $c = 2^{-2n}/(1-p_1)^2$.

The line element \eqref{solk}
describes a family of cosmological Kasner solutions, labelled by $n \in \mathbb{N}$, which includes
the case $n=2$ that was obtained in  \cite{Aniceto:2019rhg}. Note that while we can obtain 
solutions with exponents  $p_1 = p_3 = 0, p_2 = 1$ and $p_1 =p_2 =0, p_3 =1$ in this manner, we cannot
obtain a Kasner solution with exponents 
 $p_1 = 1, p_2 = p_3 = 0$. The latter case was studied in Section 9 of \cite{Aniceto:2019rhg}.

Each solution in this Kasner class is related by ``multiplicative deformation" with a family
of canonical solutions $M_a (\rho, v)$, for a given value of the exponent $2n$.
Therefore, conversely, we can now
view these canonical solutions as deformations of Kasner solutions.

Finally, would we have chosen the contour $\Gamma$ to be such that ${\tilde \tau}_a$ is in its interior, we would have obtained
the same family of cosmological Kasner solutions.

Let us now discuss conserved currents. To this end we return to $X_m = R X$, see \eqref{Ri}. Upon setting $\tau = \varphi_{\omega}$, $X_m$ not only denotes a solution to the linear
system, see Theorem \ref{theoMN},
but also constitutes the quantity that enables one to define a conserved current, called 
Kac-Moody current, given by ${\hat J}  = \star d \left( X_m^{-1} \partial_{\omega} X_m \right)$
\cite{Lu:2007jc,Nicolai:1991tt}.

 Let us exhibit the Kac-Moody current $\hat J$  associated to a Kasner solution \eqref{Mmk}. Using 
 \be
X_m (\tau, \rho, v) = R (\tau, \rho, v) X (\tau, \rho, v) = \begin{bmatrix}
	(\tau - \tau_a)^n (\tau - {\tilde \tau}_a)^n
		\; \; & \;\;	0 \\
		0\;  \; & \; \;	(\tau - \tau_a)^{-n} (\tau - {\tilde \tau}_a)^{-n}
\end{bmatrix} ,
\label{Xmk}
\ee
we obtain
\bea
 X_m^{-1}(\omega, \rho, v) \partial_{\omega} X_m (\omega, \rho, v)= n 
 \begin{bmatrix}
 1 & 0 \\ 
 0 & -1 
 \end{bmatrix} 
 \left( \frac{1}{\omega - a}
 + \frac{\partial_{\omega} \varphi_{ \omega} ( \rho, v) }{\varphi_{\omega} (\rho, v)} \right) \;,
  \eea
where $X_m (\omega, \rho, v)$ denotes the composition of $X_m(\tau, \rho, v)$ with $\tau = \varphi_{\omega}$ 
given in \eqref{vpp},
with $\lambda = -1$.
This results in
\bea
{\hat J}  = \star d \left( X_m^{-1} \partial_{\omega} X_m \right) =  n  \begin{bmatrix}
 1 & 0 \\ 
 0 & -1 
 \end{bmatrix}  \, \star d \left( \frac{\partial_{\omega} \varphi_{\omega} ( \rho, v) }{\varphi_{\omega} (\rho, v)} \right) \;.
\eea
Note that this expression is independent of $a$ in  \eqref{monkas}.
Using 
 \bea
\partial_{\omega} \varphi_{\omega} = 2  \,  \frac{\varphi_{\omega}^2}{\rho \, \left( \varphi_{\omega}^2 -1  \right) }\;,
\label{deromphi}
\eea
we obtain
\bea
 d \left( \frac{\partial_{\omega} \varphi_{\omega} ( \rho, v) }{\varphi_{\omega} (\rho, v)} \right) = 
- 4 \frac{\varphi_{\omega}^2}{\rho^2 ( \varphi_{\omega}^2 -1)^3 } \Big[ (- \varphi_{\omega}^2 -1) \, dv - 2 \varphi_{\omega} \, d \rho \Big] \;,
\eea
and hence, using \eqref{stho},
\bea
{\hat J} =   - 4  n \, \begin{bmatrix}
 1 & 0 \\ 
 0 & -1 
 \end{bmatrix}  \,
  \frac{\varphi_{\omega}^2}{\rho^2 ( \varphi_{\omega}^2 -1)^3 } \Big[ (- \varphi_{\omega}^2 -1) \, d\rho -2   \, \varphi_{\omega} \, d v \Big] \;.
\eea

\section{Einstein-Rosen waves \label{E-R}}

Let us look for solutions 
in the context of the four-dimensional Einstein field equations in vacuum. In this case,
 the involution $\natural$
is matrix transposition, and
the matrix $M(\rho,v)$ is a two-by-two matrix,
\bea
\label{M22}
M =
\begin{pmatrix}
 \Delta + {\tilde B}^2/\Delta &\;  {\tilde B}/ \Delta\\
{\tilde B}/\Delta  &\;  1/\Delta
\end{pmatrix} \;,
\eea
where $\Delta$ and $\tilde B$ are functions of the Weyl coordinates $(\rho,v)$. These functions determine the four-dimensional line element \eqref{4dWLP}
(in particular, ${\tilde B}$ determines $B$ through $ \rho \star d {\tilde B} = \Delta^2 \, dB $
\cite{Breitenlohner:1986um}, and $\psi$ is obtained by integrating \eqref{eq_psi}). 
In the example we will consider here, 
$ds_2^2 =  d \rho^2 -  dv^2 $ and $\lambda =-1$ (cf. \eqref{4dWLP}).

We consider the following family of diagonal matrices,
\bea
 {\cal M}(\omega) =
\begin{pmatrix}
e^{ 4 b \, e^{- a k}  \cos \left( k  \omega \right) }& 0 \\
0 & e^{ - 4 b \, e^{- a k}  \cos \left( k  \omega \right) }
\end{pmatrix} \;,
\label{monpesuper}
\eea
where  $a, b \in \mathbb{R}, a >0$ are constants, and where the parameter $k$ takes values in $\mathbb{R}_0^+$.
If we set $k=1$ and choose $b = \frac12 e^a$, we obtain the matrix that was
studied recently in \cite{Penna:2021kua} by a different method. 

We define the monodromy matrix by composition with the spectral relation \eqref{spec} with $\lambda =-1$,
\bea
{\cal M}_{\rho,v}   (\tau) = {\cal M} \left(  v + \frac{\rho}{2} \, \frac{1 +  \tau^2}{\tau} \right) \;.
\eea
Each diagonal element of ${\cal M}_{\rho,v} (\tau)$ has essential singularities at $\tau =0$ and $\tau = \infty$. However, the presence of these
singularities does not constitute a problem in our Riemann-Hilbert approach, since these functions are H\"older continuous, and indeed analytic, on any 
contour $\Gamma$ as considered here
(see Remark \ref{admc}).

{\remark \label{conhom}
 Note that all such contours
are homotopic in the punctured complex plane $\mathbb{C} \backslash \{0\}$, where ${\cal M}_{\rho,v}   (\tau)$ is analytic, and so we take $\Gamma$
to be any admissible contour.\\ 
}

To factorise ${\cal M}_{\rho,v}   (\tau) $ it is enough to determine a canonical Wiener-Hopf factorisation of the scalar function 
\bea
\exp\, [f ( v + \frac{\rho}{2} \, \frac{1 +  \tau^2}{\tau} )], \,\,\,\,\,{\rm with}\,\,\,f(\omega) = 4 b \, e^{- a k}  \cos  \left( k  \omega \right) , 
\eea
w.r.t. $\Gamma$, which can easily be done by additively decomposing the exponent using the projections $P^+_{\Gamma}$ and $P^-_{\Gamma}$ defined in 
Section \ref{sec:2c},
\bea
f ( v + \frac{\rho}{2} \, \frac{1 +  \tau^2}{\tau} )=P^+_{\Gamma} \left(f ( v + \frac{\rho}{2} \, \frac{1 +  \tau^2}{\tau} ) \right)
+ P^-_{\Gamma} \left(f ( v + \frac{\rho}{2} \, \frac{1 +  \tau^2}{\tau}) \right).
\eea
Thus we obtain
\bea
\exp\, [f ( v + \frac{\rho}{2} \, \frac{1 +  \tau^2}{\tau} )]=\left(e_-\,e_+(0)\right) \,\left((e_+(0))^{-1}e_+\right),
\eea
where
\bea
e_\pm=\exp \,P^\pm_{\Gamma} \left( f ( v + \frac{\rho}{2} \, \frac{1 +  \tau^2}{\tau} )\right).
\eea
Since $P^-_{\Gamma} \left(f ( v + \frac{\rho}{2} \, \frac{1 +  \tau^2}{\tau}) \right)$ tends to $0$ when $\tau$ tends to $\infty$, the limit of the ``minus" factor of the above factorisation, $e_-\,e_+(0)$, when $\tau$ tends to $\infty$, is 
\bea
e_+(0)=\exp \left[ P^\pm_{\Gamma} \left( f ( v + \frac{\rho}{2} \, \frac{1 +  \tau^2}{\tau} )\right) \right]_{\tau =0},
\eea
with
 \bea
\left[  P^+_{\Gamma} \left( f ( v + \frac{\rho}{2} \, \frac{1 +  \tau^2}{\tau} )\right)\right]_{\tau =0}
 &=& \frac{2 b \, e^{- a k} }{ \pi i} \int_{\Gamma} \, \frac{\cos \left( k \left(  v + \frac{\rho}{2} \, \left( \frac{1 + z^2}{z}    \right) \right) \right)}{z } \, dz \nonumber\\
  &=& 4 b \, e^{- a k}  \cos (k v) \,   J_0 (k \rho) \;,
 \eea
where we used the integral representation of the Bessel function of the first kind $J_0$,
 \bea
 J_0 (\rho) = \frac{1}{2 \pi i}  \int_{\Gamma} \,
 \frac{  e^{  \rho \, \left( z - \frac{1}{z}    \right)/2 } }{z} \, dz \;,
 \eea
 which satisfies $J_0 (- \rho) = J_0 (\rho)$. Then, from Theorem \ref{theohc}, 
  \bea
  \label{Mer}
 M(\rho,v) =
{\rm diag} (
e_+(0),  (e_+(0))^{-1} )
\eea
gives a solution to \eqref{fi2d} with $ \Delta=e_+(0)$.
 The associated four-dimensional solution \eqref{4dWLP} is
\bea
ds_4^2 =  \Delta \, dy^2 + \Delta^{-1}
\left(  e^\psi \, \left(  d \rho^2 -  dv^2 \right) + \rho^2 d\phi^2 \right) \;,
\label{ersol}
\eea
where
the metric factor $\psi (\rho, v)$ follows from \eqref{eq_psi} by integration,
\bea
\psi (\rho, v) = \left( 2 b \, e^{- a k} \right)^2  \left( k^2 \rho^2 J^2_0 (k \rho) + k^2 \rho^2 J^2_1 (k \rho) -
2 k \cos^2(k v)  \,   \rho \, J_0(k \rho) J_1 (k \rho) \right).
\eea
In the last equality
we have dropped an integration constant, and $J_1$ denotes a Bessel function of the first kind.

Setting $k=1$ and $b = \frac12 e^a$,
the metric \eqref{ersol} describes the Einstein-Rosen wave solution \cite{ER} recently discussed in \cite{Penna:2021kua}.

A remarkable property of solutions in this class is the following.

{\proposition 
Multiplying a solution of the form \eqref{Mer} by any solution 
with the same value $\lambda$ and the
same line element $ds_2^2$, obtained from a factorisation of a diagonal
monodromy matrix with respect to any admissible contour, we obtain a new solution to the field equation \eqref{fi2d}.

} 

\begin{proof}
This follows from Theorem \ref{theoMN}, taking into account  Remark \ref{conhom}, since both solutions can be obtained from 
the factorisation of diagonal monodromy matrices with respect to the same admissible contour $\Gamma$.

\end{proof}

As an example thereof, 
we return to the Kasner solution \eqref{solk}, which has $\lambda =-1$,
and interchange $\rho$ and $v$ in \eqref{kmrv} so as to obtain $ds_2^2 = d \rho^2 - dv^2$. 
This results in the four-dimensional line element
\bea
ds_4^2 =   dt^2 - t^{2 p_1}  \, d x_1^2 + t^{2 p_2}  \, d x_2^2 +  t^{2 p_3}  \, d x_3^2  \;.
\label{solsk}
\eea
Since now $x_1$ plays the role of a time-like variable, this exact solution of Einstein's field equations is also referred to as a static
Kasner solution \cite{Sabra:2020gio}. Now recall that the Einstein-Rosen wave solution \eqref{ersol} is also based on $ds_2^2 = d \rho^2 - dv^2$. 
Both the static Kasner solution and the Einstein-Rosen wave solution are obtained by factorisation of a diagonal monodromy matrix, as discussed above.
Recall that the monodromy matrix associated with the latter solution has essential singularities at $\tau = 0$ and $\tau = \infty$, while the monodromy
matrix and its inverse associated with the static Kasner solution have poles at $\tau =0, \tau_a, {\tilde \tau}_a = 1/\tau_a, \tau = \infty$, cf. \eqref{monkas}.
The monodromy matrices associated to both solutions can be factorised with respect to the {\it same} contour $\Gamma$ that avoids all these singularities, for instance
the contour $\Gamma_{\tau_a}$ that was used to perform the factorisation \eqref{canKas} of the Kasner monodromy matrix.
Denoting the static Kasner solution and the Einstein-Rosen wave solution with $k=1$ by $M_1 (\rho,v)$ and $M_2 (\rho,v)$, respectively, 
the product $M(\rho,v) = M_1 (\rho,v) M_2 (\rho,v)$ also yields a solution of the form \eqref{ersol} 
to Einstein's field equations, where
\bea
M(\rho, v) &=& 
\begin{pmatrix}
\Delta (\rho,v) & 0  \\
0  &   \Delta^{-1} (\rho, v)
\end{pmatrix}  \\
&=&  {\rm diag} \left( \left( \frac12 \rho \right)^{2n}, \left(\frac12 \rho \right)^{-2n} \right) \, 
{\rm diag} \left( e^{2b \, \cos v \, J_0 (\rho)},  e^{-2b \, \cos v \, J_0 (\rho)} \nonumber
\right), 
\eea
with $\psi$ determined from \eqref{eq_psi} by integration,
\bea 
\psi (\rho, v) =2 \, \frac{J_0(\rho) \left( n - b \, J_1(\rho) \cos v \right)^2 }{\rho \, J_1(\rho)} + 
\int d\rho \, J_0^2(\rho) \left( - \frac{2 n^2}{\rho \, J_1^2 (\rho)} + 2 b^2 \rho \right) .
\eea
Obtaining this rather complicated solution by directly solving the non-linear field equations would be highly non-trivial, whereas here
it arises in a straightforward manner.
Hence, we may view the solution  $M(\rho,v)$  as arising by a deformation of a Kasner solution $M_1 (\rho,v)$ through 
multiplication by $M_2 (\rho,v)$, an Einstein-Rosen wave solution.

\section{Superposition of Einstein-Rosen waves and gravitational pulse waves \label{sec:supER}}

The family
of Einstein-Rosen wave solutions \eqref{Mer} forms an Abelian group. These solutions are parametrized by a constant $k$, which can be regarded as a wave number 
\cite{10.2307/24900449,Weber:1957oib}.
When we multiply such solutions, of the form
\bea\label{ER}
{\rm diag} (
e^{{ f}_k} , e^{-{ f}_k})
\eea
 we are adding their logarithms, so we can interpret a product of two Einstein-Rosen wave solutions 
\eqref{Mer} with
different parameter $k$ as a superposition of waves. {From} Theorem \ref{theoMN} 
it follows that if we ``superpose", by multiplication, a finite number of
monodromy matrices of the form
\bea\label{ERw}
{\rm diag} (
e^{{\tilde f}_k(\omega)} , e^{-{\tilde f}_k(\omega)})
\eea
 with different values of the parameter $k$, the factorisation of the resulting monodromy matrix provides a solution that is a corresponding superposition of solutions to the equation \eqref{fi2d}.

One can therefore ask whether this result, which involves any finite number of monodromy matrices of the form \eqref{ERw}, can be extended to the superposition of  all matrices in that class, i.e., whether
by factorizing the monodromy matrix 
\bea
\label{monintk}
 {\cal M}(\omega) =
\begin{pmatrix}
e^{  \int_0^{\infty} {\tilde f}_k(\omega) dk }& 0 \\
0 & e^{ - \int_0^{\infty} {\tilde f}_k(\omega) dk }
\end{pmatrix} 
\eea
we get a solution to \eqref{fi2d} which is the corresponding superposition of solutions  obtained from \eqref{ERw}.

We proceed to show that the answer to this question is in the affirmative for the superposition of Einstein-Rosen waves with different wave numbers, which is referred to as gravitational pulse wave 
\cite{Ashtekar:1996cm, 10.2307/24900449,Weber:1957oib}, but no longer with respect to an arbitrary contour $\Gamma$.
{\em Whether, or under which conditions, this conclusion can be extended to any superposition of an infinite number of solutions 
to the field equation \eqref{fi2d} belonging to an Abelian group, is left here as an open question.}

\vskip 5mm

The integrals in \eqref{monintk} are convergent provided that 
$- a < {\rm Im} \, \omega < a$, in which case we obtain
\bea
{\cal M} (\omega) = \begin{pmatrix}
e^{  \int_0^{\infty} {\tilde f}_k(\omega) dk }& 0 \\
0 & e^{ - \int_0^{\infty} {\tilde f}_k(\omega) dk }
\end{pmatrix} 
=
\begin{pmatrix}
e^{ \frac{ 4 a b}{\omega^2 + a^2} }  & 0 \\
0 & e^{ -\frac{ 4 a b}{ \omega^2 + a^2} } 
\end{pmatrix} \;,
\eea
with the corresponding monodromy matrix given by
\bea
{\cal M}_{\rho,v} (\tau) = {\cal M} \left( v + \frac{\rho}{2} \frac{1 + \tau^2}{\tau} \right) .
\label{monpw}
\eea

We now determine the canonical Wiener-Hopf factorisation of 
the monodromy matrix 
\eqref{monpw}.
Let
\bea
f_{\rho, v} (\tau) = \frac{ 4 a b}{ \omega^2 + a^2} {\Big \vert}_{ \omega= v + \frac{\rho}{2} \, \frac{1 +  \tau^2}{\tau} } \;,
\eea
which has four poles located at
\bea
\tau_a = \frac{- (v + ia) + \sqrt{(v + i a )^2 - \rho^2}}{\rho} \;\;\;,\;\;\; {\tilde \tau} _a = \frac{- (v + ia) - \sqrt{(v + i a )^2 - \rho^2}}{\rho} \;, 
\nonumber\\
\tau_{-a} = \frac{- (v - ia) + \sqrt{(v - i a )^2 - \rho^2}}{\rho} \;\;\;,\;\;\; {\tilde \tau} _{-a} = \frac{- (v - ia) - \sqrt{(v - i a )^2 - \rho^2}}{\rho} \;,
\label{tpma}
\eea
where we take the principal branch of the square root. For any admissible
contour we
have that if $\tau_a$ is in the interior $\mathbb{D}^+_{\Gamma}$, then ${\tilde \tau}_a$ is in the exterior $\mathbb{D}^-_{\Gamma}$ and vice-versa;
analogously for $\tau_{-a}$ and ${\tilde \tau} _{-a}$ \cite{Aniceto:2019rhg}, because 
${\tilde \tau}_{\pm a} = 1/\tau_{\pm a}$. Note that although
the four points \eqref{tpma} are essential singularities of the 
diagonal elements of ${\cal M}_{\rho,v}   (\tau)$, we can choose $\Gamma$ in such a way that ${\cal M}_{\rho,v}   (\tau)$ is H\"older continuous
on $\Gamma$.

There are thus four non-homotopic choices for $\Gamma$ on $\mathbb{C} \backslash \{0, \tau_a, \tau_{-a}, {\tilde \tau}_{a}, {\tilde \tau}_{-a} \}$, 
specified by: either $\{ \tau_a, \tau_{-a} \}$, or $\{ \tau_a, {\tilde \tau}_{-a} \}$, or
$\{ \tau_{-a},  {\tilde \tau}_{a} \}$, or $\{ {\tilde \tau}_{a}, {\tilde \tau}_{-a} \}$ are contained in $\mathbb{D}^-_{\Gamma}$. Let us assume that $A, B, C, D$
are complex constants such that
\bea
f_{\rho, v} (\tau)  = \frac{A}{\tau - \tau_a} +  \frac{B}{\tau - \tau_{-a}} +  \frac{C}{\tau - {\tilde \tau}_{a}} + \frac{D}{\tau - {\tilde \tau}_{-a}} \;.
\label{ABCDdecom}
\eea
For any choice of $\Gamma$, $P^+_{\Gamma} f$ will be the sum of the two terms on the right hand side of \eqref{ABCDdecom} with poles in 
 $\mathbb{D}^-_{\Gamma}$. Thus, considering first the case where  $\{ \tau_a, {\tilde \tau}_{-a} \} \subset \mathbb{D}^-_{\Gamma}$,
 we have
\bea
\left( P^+_{\Gamma} \right) f_{\rho, v} (\tau) = \frac{A}{\tau - \tau_a} + \frac{D}{\tau - {\tilde \tau}_{-a}} \;\;\;,\;\;\; 
\left( P^+_{\Gamma} \right) f_{\rho, v} (0) = - \frac{A}{ \tau_a} - \frac{D}{ {\tilde \tau}_{-a}}  \;,
\eea
which, taking into account the expressions for $A$ and $D$, leads to 
\bea
M(\rho, v) = 
\begin{pmatrix}
 e^{ \frac{4 i b }{\rho} \left(\frac{1}{(\tau_a - {\tilde \tau}_a ) } +
\frac{1}{(\tau_{-a} -{\tilde \tau}_{-a} ) } \right)  }& 0 \\
0  &  e^{ - \frac{4 i b }{\rho} \left(\frac{1}{(\tau_a - {\tilde \tau}_a ) } +
\frac{1}{(\tau_{-a} -{\tilde \tau}_{-a} ) } \right) }
\end{pmatrix} \;.
\eea
Proceeding analogously for the other three possible choices of $\Gamma$, and taking into account that an
overall 
sign can be absorbed into the constant $b$ in the exponents, one is lead to just two different
expressions for $M(\rho,v)$, with the first diagonal element taking the form $m_1 = \exp ( 2 b g)$, where $g$ is given by
\bea
\label{decomp2}
\frac{i}{\sqrt{(v + i a )^2 - \rho^2} } + \frac{i}{\sqrt{(v - i a )^2 - \rho^2} }
  \qquad  \text{or} \qquad   \frac{i}{\sqrt{(v + i a )^2 - \rho^2} } - \frac{i}{\sqrt{(v - i a )^2 - \rho^2} }.
 \eea
Note that the first expression in \eqref{decomp2} is imaginary, while the second takes real values. Since we are interested in space-time solutions
that are real, we only retain the latter, which can be expressed as \cite{AS}
\bea
2  \int_0^{\infty} e^{-a k} \cos (k v ) J_0( k \rho) dk   
\eea
and corresponds to choosing $\Gamma$ such that  $\{ \tau_a, {\tilde \tau}_{-a} \} \subset \mathbb{D}^-_{\Gamma}$.
 Then, 
the resulting matrix $M(\rho,v)$, which satisfies the field equation \eqref{fi2d}, is given by
\bea
 M(\rho,v) &=& 
 \begin{pmatrix}
e^{ 4b  \int_0^{\infty} e^{-a k} \cos (k v ) J_0( k \rho) dk } & 0 \\
0 & e^{ - 4b  \int_0^{\infty} e^{-a k} \cos (k v ) J_0( k \rho) dk}
\end{pmatrix} \nonumber\\
&=& {\rm diag} (
\Delta (\rho, v) ,  \Delta^{-1}  (\rho, v) ) \;,
\label{grvpsol}
\eea
which indeed describes the superposition of Einstein-Rosen
waves with different wave numbers 
\cite{Ashtekar:1996cm, 10.2307/24900449,Weber:1957oib},
and is referred to as a gravitational pulse wave.
The associated four-dimensional solution \eqref{4dWLP} is
\bea
ds_4^2 =  \Delta \, dy^2 + \Delta^{-1} 
\left(  e^\psi \, \left(  d \rho^2 -  dv^2 \right) + \rho^2 d\phi^2 \right) \;,
\eea
where
the metric factor $\psi (\rho, v)$ follows from \eqref{eq_psi} by integration and reads
\bea
\psi (\rho, v) = \frac{b^2}{a^2} \left[ 1 - \frac{ 2 a^2 \rho^2 [ ( a^2 + \rho^2 - v^2)^2 - 4 a^2 v^2 ] }{ [ ( a^2 + \rho^2 - v^2)^2 + 4 a^2 v^2 ]^2 } + 
\frac{ \rho^2 - a^2 - v^2}{\sqrt{ ( a^2 + \rho^2 - v^2)^2 + 4 a^2 v^2 }}
\right] , 
\label{psipuls}
\eea
where we have adjusted the integration constant to reproduce the result given in \cite{Ashtekar:1996cm}.

\section{Relating $A$ and $X(\tau, \rho, v)$ \label{sec:AX}}

We now show that from  
 \eqref{p2RHMN} one may also deduce various interesting relations between the matrix one-form $A = M^{-1} d M$ and the 
 matrix function $X(\tau, \rho, v)$ that result from the
canonical Wiener-Hopf factorisation w.r.t. $\Gamma$. 

We consider the relation \eqref{p2RHMN} with $R, N$ replaced by the identity matrix, in which case $G_{M,X} (\tau,\rho, v) = - \star d M(\rho, v)$ and $G_{R,N}(\tau,\rho,v)=0$
(cf. \eqref{stMG}).
By multiplying the relation  \eqref{p2RHMN} by $M^{-1}$ from the left
we obtain the following two relations,
\bea
&& \partial_{\tau} X X^{-1} = \frac{\rho}{2 \lambda} \left( \frac{\lambda}{\tau} \, A_{\rho} - A_v - \left( 1 + \frac{\lambda}{\tau^2} \right) \partial_v X X^{-1} 
\right) \;, \nonumber\\
&& \partial_{\tau} X X^{-1} = \rho \left(  \frac{1}{\tau} \, A_{\rho} - \frac{1}{\tau^2} \,  \partial_v X X^{-1} +  \frac{1}{\tau} \,  \partial_{\rho} X X^{-1} \right) 
 \;,
 \label{2rel2}
\eea
where we used that $A= M^{-1} d M = A_v \, dv + A_{\rho} \, d\rho$.

Now recall that  \eqref{p2RHMN} is valid for all $\tau \in \Gamma$.
We may therefore integrate the relations \eqref{2rel2} along $\Gamma$, using that $X, X^{-1}, \partial_v X, \partial_{\rho} X$ are analytic
not only in the interior $\mathbb{D}^+_{\Gamma}$ of $\Gamma$ but also in a neighbourhood of $\Gamma$
and that 
\bea
\left( \frac{\partial X(\tau, \rho, v) }{\partial \rho} \right)\vert_{\tau =0} = 0 \;\;\;,\;\;\; 
\left( \frac{\partial X(\tau, \rho, v) }{\partial v} \right)\vert_{\tau =0} = 0 .
\label{devrX}
\eea
Integrating the first relation 
in \eqref{2rel2} gives
\bea
A_{\rho} = \Big( \partial_{\tau} \left( \partial_v X X^{-1} \right) \Big)  \vert_{\tau =0} \;,
\label{Arppm}
\eea
and integrating the second relation also gives \eqref{Arppm} because of \eqref{devrX}.

Next, we multiply \eqref{2rel2} with $1/\tau$ and again integrate along $\Gamma$. The first relation gives
\bea
A_v = - \frac{2 \lambda}{\rho} \,\Big( \partial_{\tau}X X^{-1} \Big) \vert_{\tau =0} - \frac{\lambda}{2}\,
\Big( \partial^2_{\tau} \left( \partial_v X X^{-1} \right) \Big)  \vert_{\tau =0} \;,
\label{Avrel}
\eea
while the second relation results in 
\bea
\Big( \partial_{\tau} X X^{-1} \Big)  \vert_{\tau =0} = - \frac{\rho}{2} \,\Big( \partial^2_{\tau} \left( \partial_v X X^{-1} \right) \Big) \vert_{\tau =0} 
+ \rho \, \Big( \partial_{\tau} \left( \partial_{\rho} X X^{-1} \right) \Big) \vert_{\tau =0} \;.
\eea
Comparing with \eqref{Avrel} we obtain
\bea
A_v = - 2 \lambda \left( \partial_{\tau} \left( \partial_{\rho} X X^{-1} \right) 
- \frac14 \,\partial^2_{\tau} 
\left( \partial_v X X^{-1} \right)  
\right)  \vert_{\tau =0}  \;.
\label{Avtr}
\eea
Thus, we have obtained expressions that relate the components of the one-form $A$ to derivatives of $X$.

Recall that $A$ satisfies the matrix equation $d A + A \wedge A = 0$. This equation implies further
relations involving derivatives of $X(\tau, \rho, v)$. For instance, when $A$ is a diagonal matrix one-form, the above equation becomes
$\partial_v A_{\rho} = \partial_{\rho} A_v$. Then,  applying $\partial_v$ to \eqref{Arppm} and $\partial_{\rho}$ to \eqref{Avtr} and equating
both gives a new equation that involves higher derivatives of $X(\tau, \rho, v)$ evaluated at $\tau =0$.

\vskip 2mm

\subsection*{Acknowledgements}
The authors would like to thank the Isaac Newton Institute for Mathematical Sciences, Cambridge, for support and hospitality during the programme
{\it Black holes: bridges between number theory and holographic quantum information}
where part of the work on this paper was undertaken. This work was supported by EPSRC grant no EP/R014604/1.
This work was partially
supported by FCT/Portugal through CAMGSD, IST-ID, projects UIDB/04459/2020 and UIDP/04459/2020.

\vskip 3mm


\providecommand{\href}[2]{#2}\begingroup\raggedright

\endgroup


\begin{thebibliography}{99}


\bibitem{Ablowitz:1991xb}
M.~J. Ablowitz and P.~A. Clarkson, {\em {Solitons, nonlinear evolution
  equations and inverse scattering}}, vol.~149.
\newblock 1991.




\bibitem{AS}
M. Abramowitz and I. A. Stegun, {\it {Handbook of mathematical functions}}, 
 {\em {Dover Publications, New York, 1964}}.
 
 \bibitem{AAM}
V.~M. Adukov, N.~V. Adukova, and G.~Mishuris, {\it An explicit Wiener-Hopf
  factorisation algorithm for matrix polynomials and its exact realizations
  within ExactMPF package},  {\em Proceedings of the Royal Society A:
  Mathematical, Physical and Engineering Sciences} {\bf 478} (2022), no.~2263
  20210941,
\href{https://doi.org/10.1098/rspa.2021.0941}{{\tt
  https://doi.org/10.1098/rspa.2021.0941}}].


\bibitem{Aniceto:2019rhg}
P.~Aniceto, M.~C. C\^amara, G.~L. Cardoso, and M.~Rossell\'o, {\it {Weyl
  metrics and Wiener-Hopf factorisation}},  {\em JHEP} {\bf 05} (2020) 124,
  \href{https://arxiv.org/pdf/1910.10632.pdf}{{\tt 1910.10632}}.

\bibitem{Ashtekar:1996cm}
A.~Ashtekar, J.~Bicak, and B.~G. Schmidt, {\it {Behavior of Einstein-Rosen
  waves at null infinity}},  {\em Phys. Rev. D} {\bf 55} (1997) 687--694,
  \href{https://arxiv.org/pdf/gr-qc/9608041.pdf}{{\tt gr-qc/9608041}}.


\bibitem{Belinsky:1971nt}
V.~A. Belinsky and V.~E. Zakharov, {\it {Integration of the Einstein Equations
  by the Inverse Scattering Problem Technique and the Calculation of the Exact
  Soliton Solutions}},  {\em Sov. Phys. JETP} {\bf 48} (1978) 985--994.




\bibitem{10.2307/24900449}
W.~B. Bonnor, {\it {Non-Singular Fields in General Relativity}},  {\em Journal
  of Mathematics and Mechanics} {\bf 6} (1957), no.~2 203--214.





\bibitem{Breitenlohner:1986um}
P.~Breitenlohner and D.~Maison, {\it {On the Geroch Group}},  {\em Ann. Inst.
  H. Poincare Phys. Theor.} {\bf 46} (1987) 215.

\bibitem{Breitenlohner:1987dg}
P.~Breitenlohner, D.~Maison, and G.~W. Gibbons, {\it {Four-Dimensional Black
  Holes from Kaluza-Klein Theories}},  {\em Commun. Math. Phys.} {\bf 120}
  (1988) 295.


\bibitem{Camara:2017hez}
M.~C. C\^amara, G.~L. Cardoso, T.~Mohaupt, and S.~Nampuri, {\it {A
  Riemann-Hilbert approach to rotating attractors}},  {\em JHEP} {\bf 06}
  (2017) 123, \href{https://arxiv.org/pdf/1703.10366.pdf}{{\tt 1703.10366}}.
  
\bibitem{CDR}  
 M.~C. C\^amara, C. Diogo, and L. Rodman,
  {\it {Fredholmness of Toeplitz operators and corona problems}}, {\em J. Funct. Anal.} {\bf  259}  (2010) 1273-1299. 


\bibitem{CLS}
M.~C. C\^amara, A.~B. Lebre, and F.-O. Speck, {\it {Meromorphic factorisation,
  Partial Index Estimates and Elastodynamic Diffraction Problems}},  {\em Math.
  Nachr.} {\bf 157} (1992) 291--372.

\bibitem{CAFSPFS}
M.~C. C\^amara, A.~F. dos Santos, P.~F. dos Santos, 
{\it {Lax equations, factorization and Riemann-Hilbert problems}}, {\rm Port. Math.} {\bf 64} (2007), no. 4, pp. 509-533.


\bibitem{CAFSPFS2}
M.~C. C\^amara, A.~F. dos Santos, P.~F. dos Santos, 
{\it {Matrix Riemann-Hilbert problems and factorization on Riemann surfaces}}, {\rm J. Funct. Anal.} {\bf 255} (2008), no. 1, 228-254.



\bibitem{Cardoso:2017cgi}
G.~L. Cardoso and J.~C. Serra, {\it {New gravitational solutions via a
  Riemann-Hilbert approach}},  {\em JHEP} {\bf 03} (2018) 080,
  \href{https://arxiv.org/pdf/1711.01113.pdf}{{\tt 1711.01113}}.
  
  \bibitem{Chakrabarty:2014ora}
B.~Chakrabarty and A.~Virmani, {\it {Geroch Group Description of Black Holes}},
   {\em JHEP} {\bf 11} (2014) 068,
  \href{https://arxiv.org/pdf/1408.0875.pdf}{{\tt 1408.0875}}.


  \bibitem{CG} K.~F. Clancey and I.~Gohberg, {\it {Factorisation of Matrix Functions and Singular Integral Operators}},
in {\em {Operator Theory: Advances and Applications, vol. 3, Birkh\"auser Verlag, Basel, 1981}}.





\bibitem{Emparan:2001wk}
R.~Emparan and H.~S. Reall, {\it {Generalized Weyl solutions}},  {\em Phys.
  Rev. D} {\bf 65} (2002) 084025,
  \href{https://arxiv.org/pdf/hep-th/0110258.pdf}{{\tt hep-th/0110258}}.





\bibitem{ER}
A.~Einstein and N.~Rosen, {\it {On gravitational waves}},  {\em Journal of the
  Franklin Institute} {\bf 223} (1937) 43--54.


\bibitem{Ernst:1967wx}
F.~J. Ernst, {\it {New formulation of the axially symmetric gravitational field
  problem}},  {\em Phys. Rev.} {\bf 167} (1968) 1175--1179.

\bibitem{Ernst:1967by}
F.~J. Ernst, {\it {New Formulation of the Axially Symmetric Gravitational Field
  Problem. II}},  {\em Phys. Rev.} {\bf 168} (1968) 1415--1417.

\bibitem{Figueras:2009mc}
P.~Figueras, E.~Jamsin, J.~V. Rocha, and A.~Virmani, {\it {Integrability of
  Five Dimensional Minimal Supergravity and Charged Rotating Black Holes}},
  {\em Class. Quant. Grav.} {\bf 27} (2010) 135011,
  \href{https://arxiv.org/pdf/0912.3199.pdf}{{\tt 0912.3199}}.





\bibitem{Gardner:1967wc}
C.~S. Gardner, J.~M. Greene, M.~D. Kruskal, and R.~M. Miura, {\it {Method for
  solving the Korteweg-deVries equation}},  {\em Phys. Rev. Lett.} {\bf 19}
  (1967) 1095--1097.

\bibitem{Gohberg2003FactorizationAI}
I.~Gohberg, N.~Manojlovic, and A.~F. dos Santos, {\it Factorization and
  integrable systems : summer school in Faro, Portugal, September 2000},  2003.



\bibitem{GKR}
G.~J. {Groenewald}, M.~A. {Kaashoek}, and A.~C.~M. {Ran}, {\it {Wiener-Hopf
  factorisation indices of rational matrix functions with respect to the unit
  circle in terms of realization}},  {\em  Indag. Math.} {\bf 34} (2023), no. 2, 338-356,
   \href{https://arxiv.org/pdf/2203.07821.pdf}{{\tt
  2203.07821}}.

\bibitem{Its}
A.~R. Its, {\it {The Riemann-Hilbert Problem and Integrable Systems}},  {\em
  Notices of the AMS} {\bf 50} (2003) 1389.
  
 \bibitem{Katsimpouri:2012ky}
D.~Katsimpouri, A.~Kleinschmidt, and A.~Virmani, {\it {Inverse Scattering and
  the Geroch Group}},  {\em JHEP} {\bf 02} (2013) 011,
  \href{https://arxiv.org/pdf/1211.3044.pdf}{{\tt 1211.3044}}.


\bibitem{Korotkin:2023lrg}
D.~Korotkin and H.~Samtleben, {\it {Integrability and Einstein's Equations}},
  \href{https://arxiv.org/pdf/2311.07900.pdf}{{\tt 2311.07900}}.



\bibitem{LS}
G.~Litvinchuk and I.~Spitkovsky, {\it {factorisation of Measurable Matrix
  Functions}},  in {\em {Oper. Theory Adv. Appl., vol. 25, Birkh\"auser Verlag,
  Basel, 1987. Translated from Russian by B. Luderer, with a foreword by B.
  Silbermann}}.


\bibitem{Lu:2007jc}
H.~Lu, M.~J. Perry, and C.~N. Pope, {\it {Infinite-dimensional symmetries of
  two-dimensional coset models coupled to gravity}},  {\em Nucl. Phys. B} {\bf
  806} (2009) 656--683, \href{https://arxiv.org/pdf/0712.0615.pdf}{{\tt
  0712.0615}}.

\bibitem{KisilAMR}
A.~V. Kisil, I.~D. Abrahams, G.~Mishuris, and S.~V. Rogosin, {\it The
  Wiener-Hopf technique, its generalizations and applications: constructive
  and approximate methods},  {\em Proceedings of the Royal Society A:
  Mathematical, Physical and Engineering Sciences} {\bf 477} (2021), no.~2254
  20210533,
  \href{https://doi.org/10.1098/rspa.2021.0533}{{\tt
  https://doi.org/10.1098/rspa.2021.0533}}.


\bibitem{MP}
S.~Mikhlin and S.~Pr\"ossdorf, {\it {Singular integral operators}},   {\em
  {Springer-Verlag, Berlin, 1986. Translated from German by Albrecht B\"ottcher
  and Reinhard Lehmann}}.




\bibitem{Nicolai:1991tt}
H.~Nicolai, {\it {Two-dimensional gravities and supergravities as integrable
  system}},  {\em Lect. Notes Phys.} {\bf 396} (1991) 231--273.





\bibitem{Penna:2021kua}
R.~F. Penna, {\it {Einstein\textendash{}Rosen waves and the Geroch group}},
  {\em J. Math. Phys.} {\bf 62} (2021), no.~8 082503,
  \href{https://arxiv.org/pdf/2106.13252.pdf}{{\tt 2106.13252}}.


\bibitem{PRD}
L. P. Primachuk, S. V.  Rogosin, and M. V. Dubatovskaya, {\it{On factorisation of partly non-rational $2 \times 2$ matrix-functions}},
{\rm Trans. A. Razmadze Math. Inst.} {\bf 176} (2022), no. 3, 403-410,
\href{https://rmi.tsu.ge/transactions/TRMI-volumes/176-3/v176(3)-9.pdf}{{\tt https://rmi.tsu.ge/transactions/TRMI-volumes/176-3/v176(3)-9.pdf}}.


\bibitem{Sabra:2020gio}
W.~A. Sabra, {\it {Kasner Branes with Arbitrary Signature}},  {\em Phys. Lett.
  B} {\bf 809} (2020) 135694, [\href{https://arxiv.org/pdf/2005.03953.pdf}{{\tt
  2005.03953}}].



\bibitem{Schwarz:1995af}
J.~H. Schwarz, {\it {Classical symmetries of some two-dimensional models
  coupled to gravity}},  {\em Nucl. Phys. B} {\bf 454} (1995) 427--448,
  \href{https://arxiv.org/pdf/hep-th/9506076.pdf}{{\tt hep-th/9506076}}.


\bibitem{Weber:1957oib}
J.~Weber and J.~A. Wheeler, {\it {Reality of the Cylindrical Gravitational
  Waves of Einstein and Rosen}},  {\em Rev. Mod. Phys.} {\bf 29} (1957), no.~3
  509--515.


\end{thebibliography}
\end{document}